\documentclass[12pt]{article}

\usepackage[small]{caption}
\usepackage[doublespacing]{setspace}
\usepackage[noend]{algpseudocode}
\usepackage[english]{babel}
\usepackage{url}
\usepackage[utf8x]{inputenc}
\usepackage[T1]{fontenc}
\usepackage{adjustbox, algorithm, algpseudocode, amssymb, amsmath, amsthm, animate, apacite, bbm, bm, booktabs, color, diagbox, enumerate, epsfig, epsf, float, framed, graphicx, indentfirst, listings, lscape, mathrsfs, multicol, multirow, pstricks, pst-node, psfrag, rotating, siunitx, subcaption, threeparttable, tikz, verbatim}

\usetikzlibrary{arrows,calc,tikzmark}
\usepackage[colorlinks=true,linkcolor=blue,citecolor=blue]{hyperref}%
\usepackage[round]{natbib}
\usepackage[toc,page]{appendix}
\makeatletter
\def\BState{\State\hskip-\ALG@thistlm}
\makeatother
\makeatletter
\newcounter{phase}[algorithm]
\newlength{\phaserulewidth}
\newcommand{\setphaserulewidth}{\setlength{\phaserulewidth}}

\makeatother
\setphaserulewidth{.7pt}

\newcounter{case}[algorithm]
\newlength{\caserulewidth}
\newcommand{\setcaserulewidth}{\setlength{\caserulewidth}}

\makeatother
\setcaserulewidth{.7pt}
\algdef{SE}[SUBALG]{Indent}{EndIndent}{}{\algorithmicend\ }%
\algtext*{Indent}
\algtext*{EndIndent}

\def\diag{\hbox{diag}}

\def\diag{\hbox{diag}}

\def\boxit#1{\vbox{\hrule\hbox{\vrule\kern6pt
			\vbox{\kern6pt#1\kern6pt}\kern6pt\vrule}\hrule}}

\def\bse{\begin{eqnarray*}}
	\def\ese{\end{eqnarray*}}
\def\be{\begin{eqnarray}}
\def\ee{\end{eqnarray}}
\def\bq{\begin{equation}}
\def\eq{\end{equation}}
\def\bse{\begin{eqnarray*}}
	\def\ese{\end{eqnarray*}}

\newcommand{\0}{\mathbf{0}}

\newcommand{\bA}{\mathbf{A}}

\newcommand{\bB}{\mathbf{B}}

\newcommand{\bdelta}{\bm{\delta}}

\newcommand{\bgamma}{\bm{\gamma}}
\newcommand{\bGamma}{{\bm{\Gamma}}}
\newcommand{\bH}{\mathbf{H}}
\newcommand{\bI}{\mathbf{I}}

\newcommand{\bj}{\mathbf{j}}
\newcommand{\bL}{\mathbf{L}}
\newcommand{\bLambda}{{\bm{\Lambda}}}
\newcommand{\bmu}{\bm{\mu}}
\newcommand{\bOmega}{\bm{\Omega}}
\newcommand{\bP}{\mathbf{P}}
\newcommand{\bPsi}{{\bm{\Psi}}}

\newcommand{\bR}{\mathbf{R}}

\newcommand{\bSigma}{\bm{\Sigma}}

\newcommand{\btau}{{\bm{\tau}}}

\newcommand{\bu}{\mathbf{u}}

\newcommand{\bU}{\mathbf{U}}
\newcommand{\bV}{\mathbf{V}}
\newcommand{\bW}{\mathbf{W}}
\newcommand{\bx}{\mathbf{x}}
\newcommand{\bxi}{{\bm{\xi}}}
\newcommand{\bX}{{\mathbf{X}}}
\newcommand{\by}{\mathbf{y}}
\newcommand{\bY}{\mathbf{Y}}
\newcommand{\bz}{\mathbf{z}}
\newcommand{\bZ}{\mathbf{Z}}
\newcommand{\E}{\mathbb{E}}

\newcommand{\Prob}{\text{P}}
\newcommand{\R}{\mathbb{R}}

\newcommand{\SUN}{\ensuremath{\mathcal{SUN}}}
\newcommand{\SUT}{\ensuremath{\mathcal{SUT}}}
\newcommand{\bDelta}{\boldsymbol{\Delta}}
\newcommand{\bomega}{\boldsymbol{\omega}}

\theoremstyle{plain}

\newtheorem{cor}{Corollary}
\newtheorem{prop}{Proposition}
\newtheorem{lem}{Lemma}
\theoremstyle{definition}

\newtheorem{defi}{Definition}
\newtheorem{rem}{Remark}

\renewcommand{\Prob}{\mathbb{P}}
\newcommand\NB[1]{\kern-0.4em\raisebox{-1.5ex}{$\stackrel{\big|}{\hbox{%
  \tiny\sc NB}}$}\marginpar{\sl\footnotesize\raggedright#1\hfill}\kern-0.2em}
\long\def\Note#1{\footnote{#1}\kern-0.2em\NB{cf. Note No.\,\arabic{footnote}}}
\let\phi=\varphi
\begin{document}
\baselineskip=18pt 
\begin{center} {\large{\bf Multivariate Unified Skew-$t$ Distributions And Their Properties}
}	
\end{center}

\baselineskip=12pt \vskip 1mm

\begin{center}
Kesen Wang\footnote[1]{\baselineskip=10pt Statistics Program,
King Abdullah University of Science and Technology,
Thuwal 23955-6900, Saudi Arabia\\
E-mail: kesen.wang@kaust.edu.sa, maicon.karling@kaust.edu.sa, marc.genton@kaust.edu.sa\\
This research was supported by the
King Abdullah University of Science and Technology (KAUST)
}, 
Maicon J. Karling\textcolor{blue}{$^1$}, Reinaldo B. Arellano-Valle\footnote[2]{\baselineskip=10pt Departamento de Estadística, Pontificia Universidad Catolica de Chile, Santiago, 7820436, Chile \\ E-mail: reivalle@mat.uc.cl}, and Marc G.~Genton\textcolor{blue}{$^1$}
\end{center}
\baselineskip=14pt  \centerline{\footnotesize \today} 

\begin{center}
{{\bf \small Abstract}}
\end{center}
The unified skew-$t$ (SUT) is a flexible parametric multivariate distribution that accounts for skewness and heavy tails in the data. A few of its properties can be found scattered in the literature or in a parameterization that does not follow the original one for unified skew-normal (SUN) distributions, yet a systematic study is lacking. In this work, explicit properties of the multivariate SUT distribution are presented, such as its stochastic representations, moments, SUN-scale mixture representation, linear transformation, additivity, marginal distribution, canonical form, quadratic form, conditional distribution, change of latent dimensions, Mardia measures of multivariate skewness and kurtosis, and non-identifiability issue. These results are given in a parametrization that reduces to the original SUN distribution as a sub-model, hence facilitating the use of the SUT for applications. Several models based on the SUT distribution are provided for illustration.    
\baselineskip=14pt
\vspace{.1cm}

\noindent
{\bf Some key words:} Heavy tail, Latent variable, Selection distribution, Skewness, Unified skew-normal distribution, Unified skew-$t$ distribution

\baselineskip=16pt


\section{Introduction}

Multivariate distributions beyond the classical Gaussian framework are needed to model modern datasets. To this end, \cite{ABG2006} proposed a selection approach to obtain multivariate distributions in a unified way while being very flexible in terms of controlling skewness and kurtosis features. Specifically, given two random vectors $\bU_0\in\R^m$ and $\bU_1\in\R^d$, and a subset $C\subset \R^m$, they coined the distribution of the random vector $\bZ=(\bU_1|\bU_0\in C)$ a selection distribution.  In this way, the cumulative distribution function (cdf) of $\bZ$ can be easily computed as
\begin{equation}\label{selection:cdf}
F_\bZ(\bz) =\Prob(\bU_1\leq\bz | \bU_0\in C) =\frac{\Prob(\bU_0\in C, \bU_1\leq\bz)}{\Prob(\bU_0\in C)},\quad \bz\in\R^d.
\end{equation}
In the absolutely continuous case, the probability density function (pdf) of $\bZ$ is then 
\begin{equation}\label{selection:pdf}
f_\bZ(\bz) = f_{\bU_1|\bU_0\in C}(\bz)=f_{\bU_1}(\bz)\,\frac{\Prob(\bU_0\in C|\bU_1=\bz)}{\Prob(\bU_0\in C)},\quad \bz\in\R^d,
\end{equation}
where $f_{\bU_1}(\bz)$ is the pdf of $\bU_1$. Note that the selection pdf  (\ref{selection:pdf})  may also be motivated by 
\begin{equation*}
\Prob(\bU_0\in C)=\mathbb{E}\{\Prob(\bU_0\in C|\bU_1)\}=\int_{C}\Prob(\bU_0\in C|\bU_1=\bz)f_{\bU_1}(\bz)\mbox{d}\bz.
\end{equation*}

One of the best-known examples of a selection distribution is the multivariate unified skew-normal (SUN) distribution, studied by \cite{arellano2006unification}, that can account for skewness in the data. For this, they first defined the selection random vector $\bZ=(\bU_1|\bU_0+\btau>\0)$ with $\btau\in\R^m$ being a vector of truncation parameters and assuming that $\bU_0$ and $\bU_1$ have a multivariate normal joint distribution with zero mean and positive-definite correlation matrix $\bar \bOmega^*$  ($\bar\bOmega^*>0$ hereinafter); that is,   
\begin{equation} \label{eq00}
\begin{pmatrix}
\bU_0 \\
\bU_1
\end{pmatrix} \sim {\cal N}_{m+d}\left(\begin{pmatrix}
    \0 \\
    \0
\end{pmatrix}, \bar \bOmega^*=\begin{pmatrix}
    \Bar{\bGamma} & \bDelta^\top \\
    \bDelta & \Bar{\bOmega}
\end{pmatrix}\right),
\end{equation}
where $\Bar{\bGamma}$ and $\Bar{\bOmega}$ are the correlation matrices of $\bU_0$ and $\bU_1$, respectively, 
and $\bDelta$ is the correlation matrix between $\bU_0$ and $\bU_1$. Then they defined the SUN distribution as the distribution of $\bY=\bxi+\bomega\bZ$, with $\bomega>0$ being a diagonal $d\times d$ scale matrix, denoted by $\bY\sim \SUN_{d,m}(\bxi,\bOmega,\bDelta,\btau,\bar\bGamma)$, with $\bOmega = \bomega \bar\bOmega \bomega$,  and with pdf 
\begin{align*}\label{pdfSUN}
      f_\bY(\by) & =|\bomega^{-1}|f_\bZ\{\bomega^{-1}(\by-\bxi)\}\\
      & = \phi_d(\by;\bxi, \bOmega)\,\frac{\Phi_m \left\{\btau + \bDelta^\top\Bar{\bOmega}^{-1}\bomega^{-1}(\by - \bxi);\Bar{\bGamma} - \bDelta^\top\Bar{\bOmega}^{-1}\bDelta \right\}}{\Phi_m(\btau;\Bar{\bGamma})},\quad \by\in\mathbb{R}^d,  
    \end{align*}
where  $\phi_d(\by;\bxi, \bOmega)$ is the pdf of  ${\cal N}_d(\bxi, \bOmega)$  and $\Phi_m(\cdot;\bPsi)$ is the cdf of  ${\cal N}_m(\0,\bPsi)$; and with cdf 
$$
    F_\bY(\by) =\Prob\{\bZ \leq \bomega^{-1}(\by-\bxi)\} = \frac{\Phi_{d+m}(\by_* - \bxi_*;\bOmega_*)}{\Phi_m(\btau;\bar{\bGamma})}, \quad \by\in\mathbb{R}^d,
    $$
where 
\begin{equation}\label{eq01}
\by_* = \begin{pmatrix} \btau \\ \by \end{pmatrix}, \quad
\bxi_* = \begin{pmatrix} \0 \\ \bxi\end{pmatrix},\quad\mbox{and}\quad 
\bOmega_* =  \begin{pmatrix}
    \Bar{\bGamma} &  -\bDelta^\top\bomega \\
    -\bomega\bDelta & \bOmega
\end{pmatrix}. 
\end{equation}

Another important example of a selection distribution is the multivariate unified skew-$t$ (SUT) distribution, a flexible parametric family that can account for both skewness and heavy tails in the data. We start with a formal definition of the SUT distribution from the aforementioned selection approach. For this, we denote by ${\cal T}_d(\bxi,\bOmega,\nu)$ the $d$-dimensional multivariate $t$ distribution with location vector $\bxi\in\mathbb{R}^d$, $d\times d$ dispersion matrix $\bOmega$ and degrees-of-freedom parameter $\nu>0$; also its pdf and cdf are denoted by $t_d(\cdot;\bxi, \bOmega, \nu)$ and $T_d(\cdot-\bxi;\bOmega,\nu)$, respectively.
\begin{defi}\label{defi1}
A random vector $\bY = \bxi + \bomega \bZ$, where
$\bZ = (\bU_1|\bU_0 + \btau > \0)$, with $\btau \in \mathbb{R}^m$ and
\begin{equation} \label{eq1}
\begin{pmatrix}
\bU_0 \\
\bU_1
\end{pmatrix} \sim {\cal T}_{m+d}\left(\begin{pmatrix}
    \0 \\
    \0
\end{pmatrix}, \bar \bOmega^*=\begin{pmatrix}
    \Bar{\bGamma} & \bDelta^\top \\
    \bDelta & \Bar{\bOmega}
\end{pmatrix}, \nu \right),
\end{equation}
is said to have a multivariate unified skew-$t$ (SUT) distribution with location vector $\bxi\in\mathbb{R}^d$, $d\times d$ dispersion matrix $\bOmega=\bomega\bar{\bOmega}\bomega$, $d\times m$ skewness matrix $\bDelta$, latent truncation vector $\btau\in\mathbb{R}^m$, $m\times m$ latent correlation matrix $\bar \bGamma$, and degrees-of-freedom parameter $\nu>0$.
In brief, $\bY\sim {\cal SUT}_{d,m}(\bxi,\bOmega,\bDelta,\btau,\bar\bGamma,\nu)$, where $m$ is the latent dimension.
\end{defi}

As in the SUN case, if  $\bar{\bOmega}_*>0$ then the selection approach used to define the SUT distribution allows to easily obtain the pdf and cdf of $\bY \sim {\cal SUT}_{d,m}(\bxi,\bOmega,\bDelta,\btau,\Bar{\bGamma},\nu)$ by applying (\ref{eq1}) in (\ref{selection:pdf}) and (\ref{selection:cdf}), respectively, as follows:
\begin{equation}\label{pdfSUT}
      f_\bY(\by) = t_d(\by;\bxi, \bOmega, \nu)\frac{T_m \left [\alpha_{\nu,Q_\by}^{-1/2}\left\{\btau + \bDelta^\top\Bar{\bOmega}^{-1}\bomega^{-1}(\by - \bxi) \right \};\Bar{\bGamma} - \bDelta^\top\Bar{\bOmega}^{-1}\bDelta, \nu + d \right]}{T_m(\btau;\Bar{\bGamma}, \nu)},\quad \by\in\mathbb{R}^d,  
    \end{equation}
where $\alpha_{\nu,Q_\by}=  \{\nu + Q_\by\}/(\nu + d)$ with $Q_\by = (\by - \bxi)^\top\bOmega^{-1}(\by - \bxi)$; and
\begin{equation} \label{cdf}
F_\bY(\by) = \frac{T_{d+m}(\by_* - \bxi_*;\bOmega_*, \nu)}{T_m(\btau;\bar{\bGamma}, \nu)}, \quad \by\in\mathbb{R}^d,
\end{equation} with $\by_*$, $\bxi_*$ and $\bOmega_*$ defined as in (\ref{eq01}).
In the pdf (\ref{pdfSUT}) of the SUT, the factor $\alpha_{\nu, Q_\by}$ has to be included, whereas it does not arise in the case of the SUN. This originates from the conditional pdf generator for the multivariate elliptical distributions, including the multivariate $t$ distribution as a particular case, specified in \cite{fang2018symmetric}. 

If $\nu \rightarrow \infty$ in Definition \ref{defi1}, then (\ref{eq1}) reduces to (\ref{eq00}) and,  therefore, the SUT becomes the SUN distribution. 
If instead $m=1$ in Definition \ref{defi1}, then $\bY\sim {\cal EST}_{d}(\bxi,\bOmega,\delta,\tau,\nu)$, the extended skew-$t$ (EST) distribution introduced by \cite{AG2010}, and if in addition $\nu \rightarrow \infty$, then $\bY\sim {\cal ESN}_{d}(\bxi,\bOmega,\delta,\tau)$, the extended skew-normal (ESN) distribution. Finally, if $m=1$ and $\btau=\0$ in Definition \ref{defi1}, then $\bY\sim {\cal ST}_{d}(\bxi,\bOmega,\bdelta,\nu)$, the skew-$t$ (ST) distribution in the form introduced by \cite{azzalini2003distributions}, and if in addition $\nu \rightarrow \infty$, then $\bY\sim {\cal SN}_{d}(\bxi,\bOmega,\bdelta)$, the skew-normal (SN) distribution of \cite{azza1996}. The multivariate $t$ distribution, $\bY\sim {\cal T}_d(\bxi,\bOmega,\nu)$, is recovered by  setting  $\btau=\0$ and $\bDelta=\0$ in Definition \ref{defi1}, from which we obtain the multivariate normal distribution, $\bY\sim {\cal N}_{d}(\bxi,\bOmega)$, by taking the limit when $\nu \rightarrow \infty$.

Notice that one interesting property of the pdf and cdf of the SUT is that unlike the SUN distribution, for which $\bDelta = \0$ is sufficient to recover its corresponding elliptically symmetric parent distribution, the SUT requires both $\bDelta = \0$ and $\btau = \0$. Indeed, when $\bDelta = \0$ and $\btau = \0$, the pdf of $\bY$ becomes:
\begin{align*}
    f_\bY(\by) 
      = t_d(\by;\bxi, \bOmega, \nu)\frac{T_m (\0 ;\Bar{\bGamma}, \nu + d )}{T_m(\0;\Bar{\bGamma}, \nu)} = t_d(\by;\bxi, \bOmega, \nu), \quad \by\in\mathbb{R}^d,
\end{align*}
since $T_m (\0 ;\Bar{\bGamma}, \nu + d ) = T_m(\0;\Bar{\bGamma}, \nu)=\Phi_m(\0;\bar\bGamma)$ according to the properties proved in \cite{fang2018symmetric} and detailed later in Section \ref{4.3}. Therefore, the cdf of $\bY$ is then $T_d(\by -\bxi; \bOmega, \nu)$.

\begin{figure}[t!]
\centering
\begin{subfigure}{0.3\textwidth} 
  \centering
  \includegraphics[width=1\textwidth,]{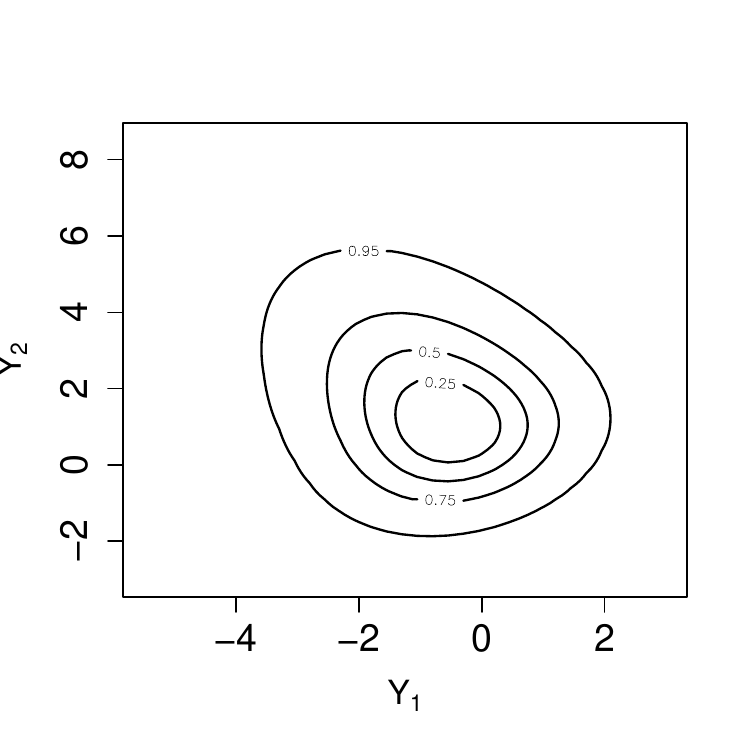}
  \caption{$\SUN_{2,1}\equiv {\cal SN}_2$}
  \label{SUN1}
\end{subfigure}
\hspace{1mm}
\begin{subfigure}{0.3\textwidth}
  \centering
  \includegraphics[width=1\textwidth,]{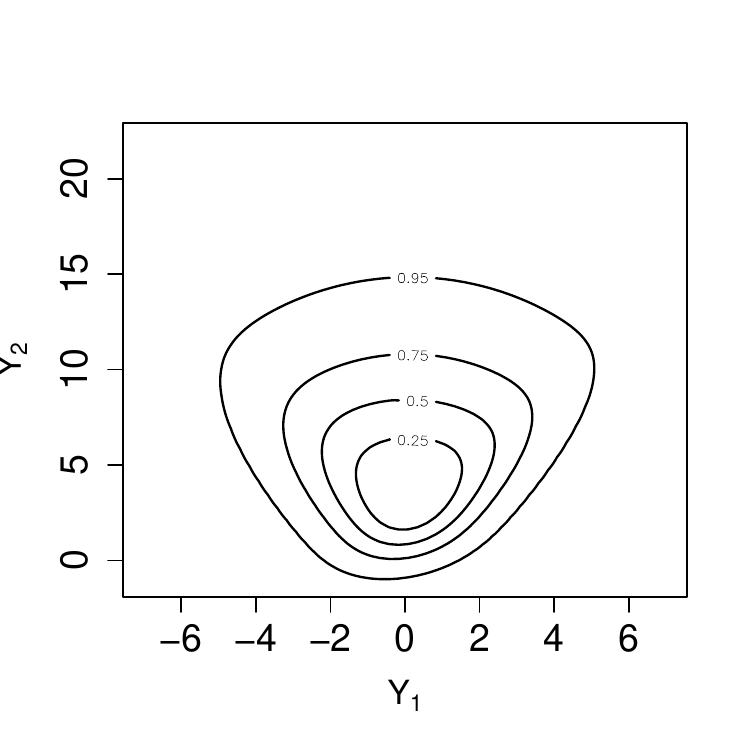}  
  \caption{$\SUN_{2,2}$}
   \label{SUN2}
\end{subfigure}
\hspace{1mm}
\vspace{1mm}
\begin{subfigure}{0.3\textwidth}
  \centering
  \includegraphics[width=1\textwidth,]{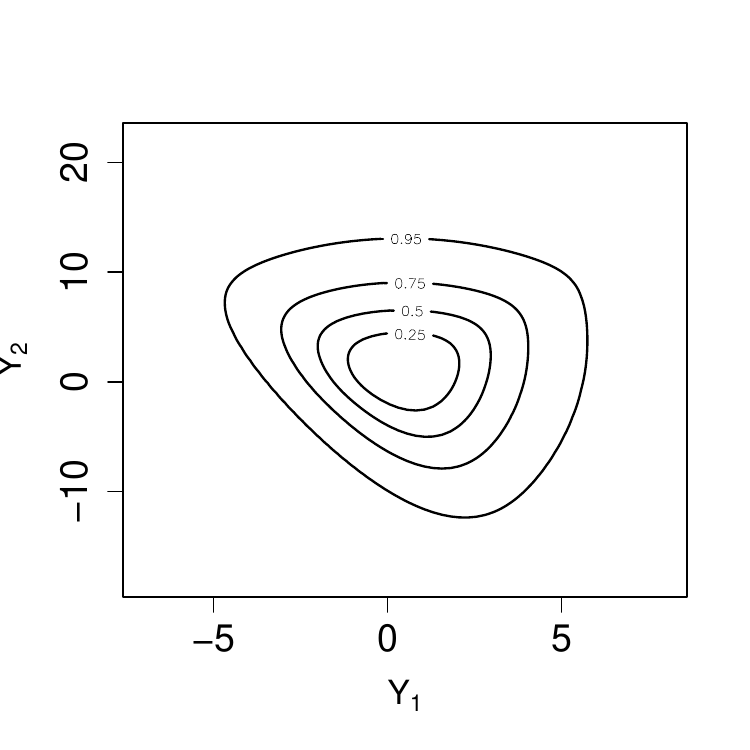}   
  \caption{$\SUN_{2,3}$}
  \label{SUN3}
\end{subfigure}
\hspace{1mm}
\vspace{1mm}
\\
\begin{subfigure}{0.3\textwidth} 
  \centering
  \includegraphics[width=1\textwidth,]{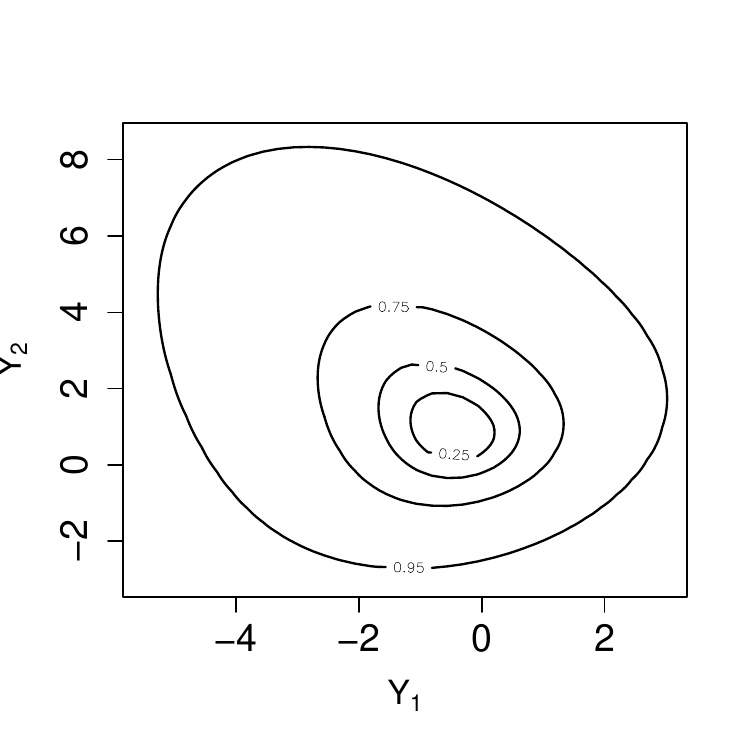}
  \caption{${\cal SUT}_{2,1}\equiv {\cal ST}_2$}
  \label{SUT1}
\end{subfigure}
\hspace{1mm}
\begin{subfigure}{0.3\textwidth}
  \centering
  \includegraphics[width=1\textwidth,]{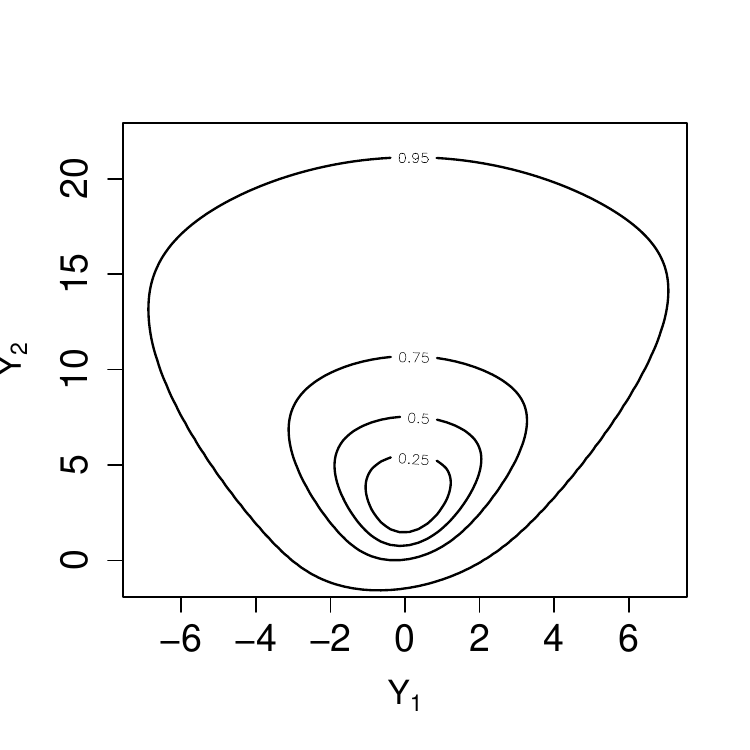}  
  \caption{${\cal SUT}_{2,2}$}
   \label{SUT2}
\end{subfigure}
\hspace{1mm}
\begin{subfigure}{0.3\textwidth}
  \centering
  \includegraphics[width=1\textwidth,]{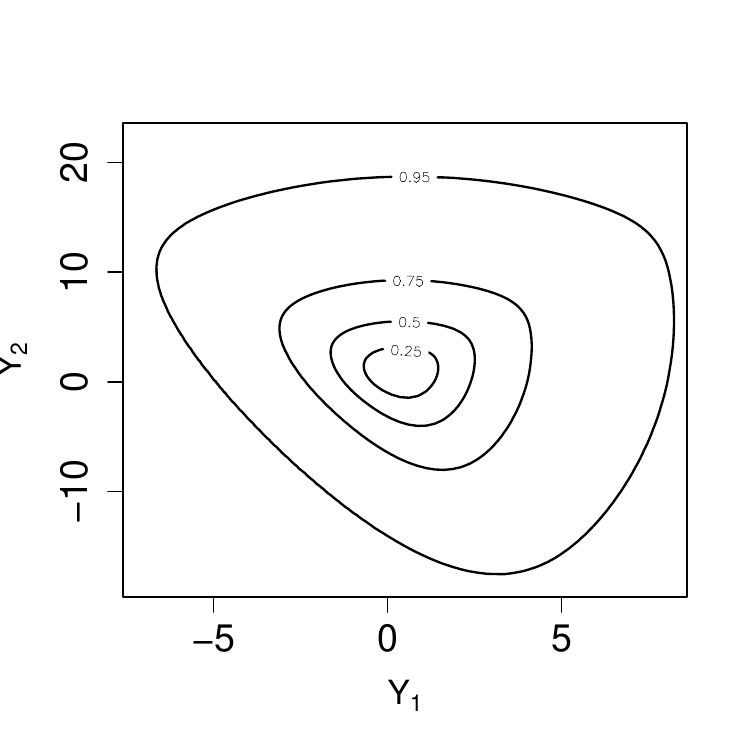}   
  \caption{${\cal SUT}_{2,3}$}
  \label{SUT3}
\end{subfigure}
\caption{Contours of the pdfs of the bivariate SUN and SUT with the same parameter specification at $\nu = 5$ with $(\bOmega, \bDelta, \bar\bGamma)$ specified to make the distribution skewed in the direction $(-1,2)^\top$ ($m=1$), $\{(-1,2)^\top,(1,2)^\top\}$ ($m=2$), and $\{(-1,2)^\top,(1,2)^\top,(1,-6)^\top\}$ ($m=3$), respectively.}
\label{contour}
\end{figure}

We present in Figure \ref{contour} the pdf contours of the bivariate SUN and SUT random vectors with the same parameter specifications at $\nu = 5$. 
Figure \ref{contour} indicates that the SUN and SUT distributions are flexible in terms of the direction of skewness. With desired specifications of $(\bOmega, \bDelta, \bar\bGamma)$, it is practical to enable the contours to be skewed simultaneously toward any possible directions as shown in Figures \ref{SUN1}-\ref{SUT3}. In particular, by the convolution representation of SUT (see Section \ref{convolution}) and of SUN \citep{arellano2006unification}, the column vectors of $\bomega\bDelta\Bar{\bGamma}^{-1}$ dictate the directions of skewness. In addition, compared with SUN, the SUT possesses more weights on the tails, which is the desirable tail-heavy property. The central bulks of the SUN, on the other hand, contain more weight than the SUT.


The formal properties of the aforementioned sub-models of Definition \ref{defi1} have been investigated in detail in the literature. For example, the properties of the SUN distribution can be found in \cite{arellano2006unification}, \cite{gupta2013}, \cite{arellano2022some}, and \cite{nonke}, whereas the properties of the EST distribution are described in \cite{AG2010}. Moreover, book-length accounts of the properties of the SN and ST distributions can be found in \cite{AC2014} and \cite{genton2004}; see also the recent review by \cite{azza2022} and references therein.

A few of the properties of the multivariate SUT distribution can be found scattered in the literature. In principle, these properties can be derived from those of the unified skew-elliptical (SUE) distributions studied by \cite{arellano2010multivariate} by plugging the Student's $t$ generator in these results, but this is a cumbersome task. Moreover, the parameterization used by \cite{arellano2010multivariate} does not follow the original one for the SUN distributions introduced by \cite{arellano2006unification}. Hence, a systematic study of the SUT distribution and its properties in the original parameterization of the SUN distribution is lacking. 

In this work, explicit properties of the multivariate SUT distribution are presented, such as its  stochastic representations, moments, SUN-scale mixture representation, linear transformation, additivity, marginal distribution, canonical form, quadratic form, conditional distribution, change of latent dimensions, Mardia measures of multivariate skewness and kurtosis, and non-identifiability issue. These results are given in the parametrization of Definition \ref{defi1} that reduces to the original SUN distribution as a sub-model, hence facilitating the use of the SUT for applications. Several models based on the SUT distribution are provided for illustration.   

The remainder of this paper is structured as follows. Section \ref{s1} discusses stochastic representation methods for the construction of the SUT with different possible parameterizations, the moments of the SUT, and relates the SUT to the SUN. Section \ref{s4} describes some of the SUT's main properties, including linear transformation, additivity, marginal distribution, canonical form, and quadratic forms. Section \ref{s5} provides the SUT's conditional distributions, as well as change of latent dimensions. Section \ref{s7} describes the Mardia measures of multivariate skewness and kurtosis of the SUT. Lastly, Section \ref{s9} discusses a non-identifiability problem of the SUT with its remedies, and some identifiable sub-models. The paper ends with a discussion in Section~\ref{s10}.

\section{Stochastic Representations and Moments} \label{s1}

\subsection{Several Selection Representations}


As discussed in \cite{ABG2006}, the same selection vector can be represented in several equivalent ways. Therefore, various equivalent representations exist for a random vector following a SUT distribution, $\bY\sim {\cal SUT}_{d,m}(\bxi,\bOmega,\bDelta,\btau,\bar\bGamma,\nu)$. In this section, we detail several possible choices. 

From Definition \ref{defi1}, we notice that $\bY=\bxi + \bZ^*$, where  $\bZ^*=\bomega \bZ = (\bomega\bU_1|\bU_0 + \btau > \0) = (\bU_1^*|\bU_0 + \btau > \0)$, where $\bU_1^* = \bomega\bU_1$. Hence, the multivariate $t$ distribution specified in Equation~(\ref{eq1}) above can be reformulated as 
\begin{equation} 
\begin{pmatrix}
\bU_0 \\
\bU_1^*
\end{pmatrix} \sim {\cal T}_{m+d}\left(\begin{pmatrix}
    \0 \\
    \0
\end{pmatrix}, \begin{pmatrix}
    \Bar{\bGamma} &  \bDelta^\top\bomega \\
    \bomega\bDelta & \bOmega
\end{pmatrix}, \nu \right). \label{eq all}
\end{equation}
In addition, the additive constants can be integrated into the multivariate distribution to arrive at a more concise form with \\
\begin{equation} \label{eq4}
\begin{pmatrix}
\tilde{\bU}_0 \\
\tilde{\bU}_1
\end{pmatrix} \sim {\cal T}_{m+d}\left(\begin{pmatrix}
    \btau\\
    \bxi
\end{pmatrix}, \begin{pmatrix}
  \Bar{\bGamma}  &  \bDelta^\top\bomega  \\
    \bomega\bDelta & \bOmega
\end{pmatrix}, \nu \right),
\end{equation}
and $\bY = (\tilde\bU_1|\tilde{\bU}_0 > \0)$. Some other formulations are also available from the SUE case in \cite{arellano2010multivariate} with various linear transformations of the multivariate $t$ distribution and the given conditions. In particular, the random vector $\bZ$ can be reformulated as $(\bU_1|\bU_0^* < \bLambda \bU_1 + \btau)$ with $\bLambda = \bDelta^\top \Bar{\bOmega}^{-1}$ and
\begin{equation} \label{eq3}
\begin{pmatrix}
\bU_0^* \\
\bU_1
\end{pmatrix} \sim {\cal T}_{m+d}\left(\begin{pmatrix}
    \0 \\
    \0
\end{pmatrix}, \begin{pmatrix}
   \Bar{\bGamma} - \bDelta^\top\Bar{\bOmega}^{-1}\bDelta  & \0 \\
    \0 & \Bar{\bOmega}
\end{pmatrix}, \nu \right).
\end{equation}
The equivalence relationship originates in the given condition $\bU_0^* < \bLambda\bU_1 + \btau$, which can also be expressed as $\bU_0^* - \bLambda\bU_1 < \btau$. Notice that \eqref{eq1} can be retrieved by setting $\bU_0 = \bLambda\bU_1 - \bU_0^*$ and applying the idempotent transformation 
$
\bA = \begin{pmatrix}
    -\bI_m & \bLambda \\
     \0 & \bI_d
\end{pmatrix} 
$ to $\begin{pmatrix}
    \bU_0^* \\
    \bU_1
\end{pmatrix}.$
Reciprocally, to reach \eqref{eq3} from \eqref{eq1}, we can set $\bU_0^* = \bLambda\bU_1 - \bU_0$ and the joint distribution $$\begin{pmatrix}
    \bU_0^* \\
    \bU_1
\end{pmatrix} = \bA \begin{pmatrix}
    \bU_0\\
    \bU_1
\end{pmatrix},$$ which still follows a multivariate $t$ distribution with mean $\0$, covariance $$ \bA \begin{pmatrix}
    \Bar{\bGamma}   & \Delta^\top \\
    \Delta & \Bar{\bOmega}
\end{pmatrix} \bA^\top = \begin{pmatrix}
   \Bar{\bGamma} - \bDelta^\top\Bar{\bOmega}^{-1}\bDelta & \0 \\
    \0 & \Bar{\bOmega} 
\end{pmatrix},
$$ and degrees-of-freedom $\nu$ as specified in \eqref{eq3}.

Another approach to arrive at the same result is by directly considering the representation $\bY = \bxi + \bomega\bZ$ where $\bZ = (\bU_1|\bU_0^* < \bLambda \bU_1 + \btau)$. Then $\bY = \bxi + (\bU_1^*|\bU_0^* < \bLambda\bomega^{-1}\bU_1^* + \btau)$. Now if we set $\bU_0 = \bLambda\bomega^{-1}\bU_1^* - \bU_0^*$, then the joint distribution of $(\bU_0^\top,{\bU_1^*}^\top)^\top$ is exactly the same as specified in \eqref{eq all} after applying the corresponding linear transformation:
$$\begin{pmatrix}
        -\bI_m & \bLambda\bomega^{-1} \\
        \0 & \bI_d
    \end{pmatrix} \text{ to } \begin{pmatrix}
        \bU_0^* \\
        \bU_1^*
    \end{pmatrix}.$$ 

Although we have summarized numerous feasible settings of the conditioning mechanism for the generation of a SUT random vector, we will only apply the setting characterized in \eqref{eq all} and \eqref{eq4} to derive the properties of the SUT in the rest of this work due to consistency and simplicity, except for its quadratic form because we will have to impose the ``uncorrelated'' assumption of $\bU_0^*$ and $\bU_1$ implied in \eqref{eq3}. 
 A summary of the constructions for selection distributions, of which the SUT is a special case, can be found in Section 4 of \cite{ABG2006}.  

\subsection{Convolution Mechanism} \label{convolution}

In addition to the selection representation, there is an equivalent convolution representation. Analogous to the SUN distribution, the SUT can be represented by the convolution of a multivariate $t$ random vector and a truncated multivariate $t$ random vector. 
\begin{prop} \label{cov}
    Let $\bY \sim {\cal SUT}_{d,m}(\bxi,\bOmega,\bDelta,\btau,\Bar{\bGamma},\nu)$. Then $\bY = \bxi + \bomega \bZ$, where $\bZ = \bDelta\Bar{\bGamma}^{-1}\bU_* + \sqrt{\frac{\nu + Q_{\bU_*}}{\nu + m}}\bW_{*}$ with $Q_{\bU_*} = \bU_*^\top \Bar{\bGamma}^{-1} \bU_*$, $\bU_* = (\bU_0|\bU_0 + \btau > \0)$, $\bU_0\sim {\cal T}_m(\0,\bar\bGamma,\nu)$, $\bW_* \sim {\cal T}_d(\0,\Bar{\bOmega} - \bDelta \Bar{\bGamma}^{-1}\bDelta^\top, \nu + m)$, and $\bU_0$ and $\bW_*$ are independent random vectors. 
\end{prop}
\begin{proof}
    The proof for this result is well-documented in equation (9) of \cite{AG2010}.
\end{proof}

\subsection{Mean and Variance} \label{mean_var}

With the convolution mechanism of the SUT, it is feasible to construct semi-explicit forms of the moments. To start, we propose the following lemma.
\begin{lem} \label{lemma1}
Let  $Q_{\bU_*} = \bU_*^\top \Bar{\bGamma}^{-1} \bU_*$, $\bU_* = (\bU_0|\bU_0 + \btau > \0)$, and $\bU_0\sim {\cal T}_m(\0,\bar\bGamma,\nu)$. Then:
\begin{equation*}
\mathbb{E}\left\{\left(\frac{\nu+Q_{\bU_*}}{\nu+m}\right)^{k/2} h(\bU_*)\right\}
=\frac{M_k(\btau,\bar\bGamma,\nu)}{T_m(\btau;\0,\bar\bGamma,\nu)} \left(\frac{\nu}{\nu+m}\right)^{k/2}   \frac{c(\nu,m)}{c(\nu-k,m)}, 
\end{equation*}
where 
\begin{align*}
c(\upsilon,r)& =\frac{\Gamma\{(\upsilon+r)/2\}}{\Gamma(\upsilon/2)(\pi\upsilon)^{r/2}},\quad \upsilon, r>0, \\
 M_k(\btau,\bar\bGamma,\nu) &= \int_{\bu > \0} h\left(\bu\right) t_m\left(\sqrt{\frac{\nu-k}{\nu}}\,\bu;\btau,\bar\bGamma,\nu-k\right)d\bu\\
&=\left(\frac{\nu}{\nu - k}\right)^{m/2}T_m(\btau;\0,\bar\bGamma,\nu-k)\mathbb{E
}\left\{h\left(\sqrt{\frac{\nu}{\nu-k}}\,\bU_k\right)\right\},
\end{align*}
with $\bU_k=(\bW_k|\bW_k>\0)$ and   $\bW_k\sim\mathcal{T}_m(\btau, \bar\bGamma,\nu-k)$ for $0<k<\nu.$
\end{lem}
\begin{proof} 
First of all, we use the result that
\[\left(\frac{\nu+Q_{\bU_*}}{\nu+m}\right)^{k/2}t_m(\bu;\bar\bGamma,\nu)=\left(\frac{\nu}{\nu+m}\right)^{k/2}\frac{c(\nu,m)}{c(\nu-k,m)}\,
t_m\left(\sqrt{\frac{\nu-k}{\nu}}\bu;\btau,\bar\bGamma,\nu-k\right),
\]
where $\nu > k >0.$
Next, with the fact that $\bU_*$ has  pdf $g(\bu)=
t_m(\bu;\btau,\bar\bGamma,\nu)/
T_m(\btau;\0,\bar\bGamma,\nu)$  for 
$\bu > \0$, we have: 
\begin{align*}
\mathbb{E}\left\{\left(\frac{\nu+Q_{\bU_*}}{\nu+m}\right)^{k/2}h(\bU_*)\right\} =& \int_{\bu>\0}\left(\frac{\nu+Q_{\bU_*}}{\nu+m}\right)^{k/2} h(\bu)g(\bu) \mbox{d}\bu\\
=&\frac{1}{T_m(\btau;\0,\bar\bGamma,\nu)}\,\int_{\bu>\0}\left(\frac{\nu+Q_{\bU_*}}{\nu+m}\right)^{k/2} h(\bu) t_m(\bu;\btau,\bar\bGamma,\nu) \mbox{d}\bu\\
=&\frac{1}{T_m(\btau;\0,\bar\bGamma,\nu)}\left(\frac{\nu}{\nu+m}\right)^{k/2}\frac{c(\nu,m)}{c(\nu-k,m)}\\
&\times\int_{\bu>\0} h(\bu) t_m\left(\sqrt{\frac{\nu-k}{\nu}}\bu;\btau,\bar\bGamma,\nu-k\right) \mbox{d}\bu.
\end{align*}
Thus, the proof is done by the change of variable $\mathbf{v}=\sqrt{\frac{\nu-k}{\nu}}\bu$ and using that the truncated vector $\bU_k$ has pdf given by $g(\bu)=t_m(\bu;\btau,\bar\bGamma,\nu-k)/T_m(\btau;\0,\bar\bGamma;\nu-k)$ for $\bu>\0$ and $\nu-k>0$.
\end{proof}

Some specific results that emerge from Lemma \ref{lemma1} are the following:\\
\noindent 1. If $h(\bu)=1$ then $M_k(\btau,\bar\bGamma,\nu)=
\left(\frac{\nu}{\nu - k}\right)^{m/2} T_m(\btau;\0,\bar\bGamma,\nu - k)$ and so\\
\begin{align*}
    \mathbb{E}\left\{\left(\frac{\nu+Q_{\bU_*}}{\nu+m}\right)^{k/2}\right\} & = \frac{ T_m(\btau;\0,\bar\bGamma,\nu - k)}{T_m(\btau;\0,\bar\bGamma,\nu)}\left(\frac{\nu}{\nu+m}\right)^{k/2}\frac{c(\nu,m)}{c(\nu-k,m)}\left(\frac{\nu}{\nu - k}\right)^{m/2}.
\end{align*}
In particular, for $k=2$: 
\begin{align}
    \eta(Q_{\bU_*})&=\mathbb{E}\left( \frac{\nu + Q_{\bU_*}}{\nu + m}\right)
 = \frac{T_m(\btau;\0,\bar\bGamma,\nu - 2)}{T_m(\btau;\0,\bar\bGamma,\nu)} \left(\frac{\nu}{\nu-2}\right) \left(\frac{\nu + m - 2}{\nu + m}\right)\label{etaQ}\\
\Longrightarrow
\mathbb{E}\left(Q_{\bU_*}\right) & = \frac{T_m(\btau;\0,\bar\bGamma,\nu - 2)}{T_m(\btau;\0,\bar\bGamma,\nu)} \frac{m \nu }{\nu - 2}.\label{EQ}
\end{align}
The result in (\ref{EQ}) gives the expectation of the quadratic form of the truncated multivariate $t$ random vector.

\noindent 2. If $k=2$ and $h(\bu)=\bu$, then
\begin{align*}
      \mathbb{E}\left\{\left(\frac{\nu+Q_{\bU_*}}{\nu+m}\right)\bU_*\right\}
     = \frac{T_m(\btau;\0,\bar\bGamma,\nu - 2)}{T_m(\btau;\0,\bar\bGamma,\nu)} \left(\frac{\nu}{\nu-2}\right)^{3/2} \left(\frac{\nu + m - 2}{\nu + m}\right)\mathbb{E}\left(\bU_k\right).
   \end{align*} 
\noindent 3. Similarly, if $k=2$ and  $h(\bu)=\bu\bu^\top$, then
\begin{align*}
    \mathbb{E}\left\{\left(\frac{\nu+Q_{\bU_*}}{\nu+m}\right) \bU_*\bU_*^\top\right\}
    & = \frac{T_m(\btau;\0,\bar\bGamma,\nu - 2)}{T_m(\btau;\0,\bar\bGamma,\nu)} \left(\frac{\nu}{\nu-2}\right)^{2} \left(\frac{\nu + m - 2}{\nu + m}\right)\mathbb{E}\left(\bU_k\bU_k^\top\right).
   \end{align*} 

Now Lemma \ref{lemma1} and these specific cases 
enable the derivations of the semi-explicit forms of the mean and variance for the SUT, as well as the expectations $\mathbb{E}(V_*)$, $\mathbb{E}(V_*\bU_*)$ and $\mathbb{E}(V_*\bU_*\bU_*^\top)$, where $V_*=\frac{\nu+Q_{U_*}}{\nu+m}$, required in Subsection \ref{con-based compute}.
\begin{prop} \label{mean}
    Let $\bY \sim {\cal SUT}_{d,m}(\bxi,\bOmega,\bDelta,\btau,\bar\bGamma,\nu)$. Then:
   \begin{align*}    
    \mathbb{E}(\bY) &= \bxi + \bomega \bDelta \Bar{\bGamma}^{-1}\mathbb{E}(\bU_*), \quad \nu>1,\\ \text{Var}(\bY) &= \bomega\{\bDelta\Bar{\bGamma}^{-1}\text{Var}(\bU_*)\Bar{\bGamma}^{-1}\bDelta^\top + \eta(Q_{\bU_*})(\bar{\bOmega} - \bDelta\Bar{\bGamma}\bDelta^\top)\}\bomega, \quad \nu>2,
    \end{align*}
     where $Q_{\bU_*}$ is the same as in Proposition \ref{cov} and $\eta(Q_{\bU_*})$ is given in  
     (\ref{etaQ}).
\end{prop}
\begin{proof}
    Following from Proposition \ref{cov}, $\bY = \bxi + \bomega\bZ$, where $\bZ = \bDelta \Bar{\bGamma}^{-1} \bU_* + \sqrt{\frac{\nu + Q_{\bU_*}}{\nu + m}} \bW_*$. Then, $\mathbb{E}(\bY) = \mathbb{E}\left\{\bxi + \bomega\left(\bDelta \Bar{\bGamma}^{-1} \bU_* + \sqrt{\frac{\nu + Q_{\bU_*}}{\nu + m}} \bW_* \right)\right\} = \bxi + \bomega \bDelta \Bar{\bGamma}^{-1}\mathbb{E}(\bU_*)$ if $\nu>1$, indicating that $\mathbb{E}(\bZ) = \bomega\bDelta\Bar{\bGamma}^{-1}\mathbb{E}(\bU_*).$
    The variance is
    $\text{Var}(\bY) = \text{Var}(\bomega\bZ) = \bomega\text{Var}(\bZ)\bomega$, where 
    \begin{align*}
    \text{Var}(\bZ)  =& \bDelta\Bar{\bGamma}^{-1}\text{Var}(\bU_*)\Bar{\bGamma}^{-1}\bDelta^\top + \text{Var}\left( \sqrt{\frac{\nu + Q_{\bU_*}}{\nu + m}} \bW_* \right) \\
     & + \bDelta \Bar{\bGamma}^{-1}\text{Cov}\left(\bU_*,\sqrt{\frac{\nu + Q_{\bU_*}}{\nu + m}}\bW_* \right) + \text{Cov}\left(\sqrt{\frac{\nu + Q_{\bU_*}}{\nu + m}}\bW_*, \bU_* \right)\Bar{\bGamma} \bDelta^\top.
    \end{align*}
    Here, 
    \begin{align*}
    \text{Cov}\left(\bU_*, \sqrt{\frac{\nu + Q_{\bU_*}}{\nu + m}}\bW_* \right) & = \mathbb{E}\left(\bU_*\sqrt{\frac{\nu + Q_{\bU_*}}{\nu + m}}\bW_*^\top \right) - \mathbb{E}\left( \bU_* \right)\mathbb{E}\left(\sqrt{\frac{\nu + Q_{\bU_*}}{\nu + m}}\bW_* \right)^\top \\
     & = \mathbb{E}\left( \bU_* \sqrt{\frac{\nu + Q_{\bU_*}}{\nu + m}}\right)\mathbb{E}(\bW_*)^\top - \mathbb{E}(\bU_*)\mathbb{E}\left( \sqrt{\frac{\nu + Q_{\bU_*}}{\nu + m}} \right) \mathbb{E}(\bW_*)^\top \\
     & = \0.
     \end{align*}
    \text{Now, if $\nu>2$, we then have} 
    \begin{align*}
    \text{Var}\left( \sqrt{\frac{\nu + Q_{\bU_*}}{\nu + m}} \bW_* \right) & = \mathbb{E}\left(\frac{\nu + Q_{\bU_*}}{\nu + m}\bW_*\bW_*^\top\right) - \mathbb{E}\left( \sqrt{\frac{\nu + Q_{\bU_*}}{\nu + m}} \bW_* \right)\mathbb{E}\left( \sqrt{\frac{\nu + Q_{\bU_*}}{\nu + m}} \bW_* \right)^\top \\
    & = \mathbb{E}\left(\frac{\nu + Q_{\bU_*}}{\nu + m}\right) \mathbb{E}(\bW_*\bW_*^\top) = \eta(Q_{\bU_*})(\Bar{\bOmega} - \bDelta\Bar{\bGamma}^{-1}\bDelta^\top),
    \end{align*}
 where $\eta(Q_{\bU_*})=\mathbb{E}\left(\frac{\nu + Q_{\bU_*}}{\nu + m}\right)$ is computed by letting $k=2$ in Lemma \ref{lemma1} as indicated in (\ref{etaQ}).
\end{proof}
 The mean vector $\mathbb{E}(\bU_*)$ and covariance matrix $\text{Var}(\bU_*)$ for the truncated multivariate $t$ distribution of $\bU_*=(\bU_0|\bU_0+\btau>\0)$, with $\bU_0\sim \mathcal{T}_m(\0,\bar\bGamma,\nu)$, can be computed numerically using methods similar to  \cite{castro2010bayesian}.
Two \textsf{R} packages allow for computations of the first and second moments of the SUT random vector, \textit{mnormt} \citep{R-mnormt} and \textit{MomTrunc} \citep{R-mtrunc}.

\subsection{Higher-Order Moments} \label{con-based compute}

Explicit moments of the non-shifted SUT random vector, $\bomega\bZ$, can be computed in the same way that the moments for the SUN are computed, which is detailed in Proposition~A.4 in the Appendix of \cite{arellano2022some}. In particular, we need to set in that Proposition~A.4:
$$\bX=\bomega\bZ,~~\bA=\bomega\bDelta\bar\bGamma^{-1},~~\bU=\bU_*,~~\bB=\bomega,~~\bV= V_*^{1/2}\bW_*,~~V_*=\frac{\nu+Q_{\bU_*}}{\nu+m},$$
where $(\bU_*,V_*)$ and $\bW_*$ are independent. In this way,  although in our case the vectors $\bU=\bU_*$ and $\bV=V_*^{1/2}\,\bW_*$ are not independent since $V_*$ depends on $\bU_*$ through $Q_{\bU_*}$, the moments up to the fourth order can be calculated semi-explicitly using our Lemma \ref{lemma1}; see directly the proof of Proposition A.4 in \cite{arellano2022some}.

We calculate only the third and fourth moments because the first and second can be obtained directly through the mean and variance computed in Section \ref{mean_var}.  Here are the detailed calculations: \vspace{-.4cm}
 \begin{align*}
     \mathbb{E}(\bX \otimes \bX\bX^\top)  =& (\bA \otimes \bA)\mu_3(\bU)\bA^\top + (\bI_{d^2} + \textbf{K}_d)(\bA \otimes \bB)\left\{\mathbb{E}(V_*\bU) \otimes \mu_2(\bW_*)\right\} \bB^\top \\
     & + (\bB \otimes \bB)\text{vec}\{\mu_2(\bW_*)\}\mathbb{E}(V_*\bU^\top)\bA^\top,
   \end{align*}
    \begin{align*}
     \mathbb{E}(\bX\bX^\top \otimes \bX\bX^\top)  =& (\bA \otimes \bA)\mu_4(\bU)(\bA \otimes \bA)^\top + (\bA \otimes \bB)\{\mathbb{E}(V_*\bU\bU^\top) \otimes \mu_2(\bW_*)\}(\bA^\top \otimes \bB^\top)\\
     & + (\bA \otimes \bB)\{\mathbb{E}(V_*\bU\bU^\top) \otimes \mu_2(\bW_*)\}(\bA^\top \otimes \bB^\top)\textbf{K}_d \\ 
     & + \textbf{K}_d(\bA \otimes \bB)\{\mathbb{E}(V_*\bU\bU^\top) \otimes \mu_2(\bW_*)\}(\bA^\top \otimes \bB^\top) \\
     & + \textbf{K}_d(\bA \otimes \bB)\{\mathbb{E}(V_*\bU\bU^\top) \otimes \mu_2(\bW_*)\}(\bA^\top \otimes \bB^\top)\textbf{K}_d \\
     & + (\bA \otimes \bA)\text{vec}\{\mathbb{E}(V_*\bU\bU^\top)\}\text{vec}\{\mu_2(\bW_*)\}^\top(\bB^\top \otimes \bB^\top) \\
     & + (\bB \otimes \bB)\text{vec}\{\mu_2(\bW_*)\}\text{vec}\{\mathbb{E}(V_*\bU\bU^\top)\}^\top (\bA^\top \otimes \bA^\top) \\
     & + (\bB \otimes \bB)\mu_2(V_*)\mu_4(\bW_*)(\bB^\top \otimes \bB^\top),
 \end{align*}
 where $\mu_k(\cdot)$ represents the $k$-th multivariate moment of the indicated variable.
 
 In the above computations, for those expectations involving the function $V_*$ of the truncated vector $\bU=\bU_*$, we must first use Lemma \ref{lemma1}  and then the procedures proposed in the literature for the calculation of truncated moments under the multivariate $t$ distribution (see specific cases 1, 2 and 3 after Lemma \ref{lemma1}). The moments that involve the variable $\bV=\bW_*$ are obtained in a similar way to what has been indicated in Section \ref{s3} with $\bW_*=V^{-1/2}\bZ_0$, where $V\sim {\cal G}amma(\nu/2,
 \nu/2)$ and $\bZ_0\sim \mathcal{N}_d(\0,\bOmega)$, and they are independent.

For the shifted SUT random vector, $\bY$, the detailed calculations of the moments are provided in Equations (A.5)-(A.8) in the Appendix of \cite{arellano2022some}.

\subsection{SUN-Scale Mixtures} \label{s3}
\subsubsection{SUN-Scale Mixture Representation}\label{s31}
\cite{gupta2003multivariate} has shown that the skew-$t$ (ST) distribution can be represented as a skew-normal (SN) scale mixture. Here, we show that the SUT distribution has a similar scale mixture representation concerning the SUN distribution under the condition that the vector of truncation parameters $\btau=\0$. 

\begin{prop} \label{p3}
    Let $\bY \sim {\cal SUT}_{d,m}(\bxi,\bOmega,\bDelta,\0,\Bar{\bGamma}, \nu)$. Then, 
    \begin{equation}\label{MixSUN}
        \bY = \bxi + V^{-1/2} \bZ_0,
    \end{equation}
    where $V \sim {\cal G}amma(\nu/2,\nu/2)$ and $\bZ_0 \sim \SUN_{d,m}(\0,\bOmega,\bDelta,\0,\Bar{\bGamma})$ are independent. 
\end{prop}

\begin{proof}
Similar to the SUT case, we refer to the stochastic representation for the SUN random vector, $\bZ_0 \overset{d}{=} (\bomega\bU_1|\bU_0  > \0)$, where according to the definition of SUN distribution from \cite{arellano2006unification}, we have that $(\bU_0^\top,\bU_1^\top)^\top$ follows the distribution in \eqref{eq00}.
Then:
\begin{align}
    \bxi + V^{-1/2}\bZ_0 & = \bxi + V^{-1/2}\left(\bomega\bU_1|\bU_0 > \0 \right) \nonumber\\
    & = \bxi + \left(V^{-1/2}\bomega\bU_1|\bU_0 > \0\right)  \nonumber \\
    & =  \bxi + (V^{-1/2}\tilde{\bU}|V^{-1/2}\bU_0 > \0 ) \label{pf3},
\end{align}
where in \eqref{pf3} we applied the independence assumption between $V$ and $\bZ_0$; the independence assumption is extended to $(V,\tilde{\bU})$ and $(V,\bU_0 )$ for conditioning mechanism. Also, we used the fact that $V^{-1/2} > 0$ with probability 1. Lastly, we have
$$
V^{-1/2} \begin{pmatrix}
    \bU_0 \\
    \tilde{\bU}
\end{pmatrix} 
\sim {\cal T}_{m+d}\left(\begin{pmatrix}
    \0 \\
    \0
\end{pmatrix},\begin{pmatrix}
    \Bar{\bGamma}  &  \bDelta^\top \bomega\\
    \bomega \bDelta  & \bOmega
\end{pmatrix}, \nu \right).$$
Therefore, \eqref{pf3} yields a SUT random vector $\bY$ according to the stochastic representation \eqref{eq all} in Section \ref{s1}.
\end{proof}

An immediate consequence of the SUN-scale mixture property is that it enables retrieval of the SUN random vector through conditioning.
\begin{cor}\label{cor1}
   Let $\bY \sim {\cal SUT}_{d,m}(\bxi,\bOmega,\bDelta,\0,\Bar{\bGamma}, \nu)$ and consider the SUN-scale representation presented in Proposition \ref{p3}. Then, for any $v > 0$, 
   \begin{equation}
       (\bY|V= v) \sim \SUN_{d,m}\left(\bxi,v^{-1}\bOmega, v^{-1/2}\bDelta,\0,\Bar{\bGamma} \right).
   \end{equation}
\end{cor}
\begin{proof}
Using the representation in \eqref{MixSUN}, we have\\
\begin{align*}
    (\bY|V = v) = \left( \bxi + V^{-1/2}\bZ_0|V = v \right) = \bxi +  v^{-1/2}\bZ_0,
\end{align*}
which follows a $\SUN_{d,m}\left(\bxi,v^{-1}\bOmega,v^{-1/2}\bDelta,\0,\Bar{\bGamma} \right)$ distribution. 
\end{proof}
Finally, we would like to remark that,  according to the selection approach used in the definition of the SUT, the restriction $\btau = \0$ is imperative in Proposition \ref{p3} and Corollary \ref{cor1} because we need  the random vector, $V^{-1/2}\bU_0$ as indicated in \eqref{pf3}, in the given condition of the selection representation to formulate a multivariate $t$ distribution. If $\btau \neq \0$,  we would have a multivariate $t$ random vector plus a scaled Gamma random vector, which is inconsistent with the setting of the selection representation of a SUT distribution.


\subsubsection{SUN-Scale Mixture-based Methods for the Moments} \label{6.2}

Semi-explicit moments can be derived through the SUN-scale mixture representation obtained in Subsection \ref{s31}. In particular, we consider $\bY =  V^{-1/2}\bZ_0 \sim {\cal SUT}_{d,m}(\0, \bOmega, \bDelta, \0, \Bar{\bGamma}, \nu)$, where $V \sim {\cal G}amma(\nu/2,\nu/2)$ and $\bZ_0 \sim \SUN_{d,m}(\0,\bOmega,\bDelta,\0,\Bar{\bGamma})$ are independent. Then,
\begin{align*}
    \mu_1(\bY) = \mathbb{E}(\bY) & = \mathbb{E}(V^{-1/2}\bZ_0) = \mathbb{E}(V^{-1/2})\mathbb{E}(\bZ_0)  = M_1\mu_1(\bZ_0).
\end{align*}
Here $\mu_i(\bZ_0)$, $i = 1,\dots,4$, indicate the multivariate moments of the SUN random vector $\bZ_0$ with explicit forms up to the fourth order as computed in \cite{arellano2022some}. The existence and computation of $M_1= \mathbb{E}(V^{-1/2})$ and the corresponding higher order moments $M_2, M_3$, and $M_4$ are well known; see, e.g., \cite{RS2001}, p. 212. Let  $X \sim {\cal G}amma(\alpha,\beta)$. Then, the moments for $X$ are $\mathbb{E}(X^k) =\beta^k \Gamma(k+ \alpha)/\Gamma(\alpha)$, $k+\alpha>0$.

The variance of the SUT random vector $\bY$ can be constructed through the second moments:
\begin{align*}
    \mu_2(\bY) = \mathbb{E}(\bY\bY^\top) = \mathbb{E}(V^{-1}\bZ_0\bZ_0^\top) = \mathbb{E}(V^{-1})\mathbb{E}(\bZ_0\bZ_0^\top) = M_2 \mu_2(\bZ_0).
\end{align*}
Therefore, $\text{Var}(\bY) = \mu_2(\bY) - \mu_1(\bY) \mu_1(\bY)^\top$. Higher-order moments of $\bY$ can also be obtained using similar approaches. For instance, the third moment is
\begin{align*}
    \mu_3(\bY) = \mathbb{E}(\bY \otimes \bY\bY^\top) = \mathbb{E}(V^{-3/2}\bZ_0 \otimes \bZ_0\bZ_0^\top) = \mathbb{E}(V^{-3/2}) \mathbb{E}(\bZ_0 \otimes \bZ_0\bZ_0^\top) = M_3 \mu_3(\bZ_0),
\end{align*}
and the fourth moment is
\begin{align*}
    \mu_4(\bY) = \mathbb{E}(\bY\bY^\top \otimes \bY\bY^\top) = \mathbb{E}(V^{-2}\bZ_0\bZ_0^\top \otimes \bZ_0\bZ_0^\top) = \mathbb{E}(V^{-2}) \mathbb{E}(\bZ_0\bZ_0^\top \otimes \bZ_0\bZ_0^\top) = M_4 \mu_4(\bZ_0).
\end{align*} 
For all of the computed moments, we set $\bxi = \0$ to avoid the tedious polynomial expansions involved in the higher-order moments so that we can lay down the foundations for the moments of the shifted random vector. Therefore, the condition $\bxi = \0$ is not a restriction. The moments can be exactly computed by recognizing that the shifted random vector is $\tilde{\bY} = \bxi + \bY \sim {\cal SUT}_{d,m}(\bxi,\bOmega,\bDelta,\0,\Bar{\bGamma})$ and by applying the expansions used in Equations (A.5)-(A.8) in \cite{arellano2022some} together with the moments computed above.

\section{Linear Transformations and Quadratic Forms} \label{s4}

\subsection{Linear Transformations} \label{4.1}

To start, we describe the properties of linear transformations of the SUT random vector.  
\begin{prop}\label{proplinear}
Let $\bY \sim {\cal SUT}_{d,m}(\bxi,\bOmega,\bDelta,\btau,\Bar{\bGamma},\nu)$. The linear transformation $\bY_{\bA} = \bA\bY + \textbf{b} $, where $\bA \in \R^{n \times d}$ with rank $n\leq d$ and $\textbf{b} \in \R^{n}$, follows a ${\cal SUT}_{n,m}(\bxi_\bA,\bOmega_\bA, \bDelta_\bA,\btau,\Bar{\bGamma},\nu)$ with $\bxi_\bA = \bA\bxi + \textbf{b}$, $\bOmega_\bA = \bA\bOmega\bA^\top$, and $\bDelta_\bA = \bomega_\bA^{-1}\bA\bomega\bDelta$, where $\bomega_\bA = \text{diag}(\bOmega_\bA)^{1/2}$.
\end{prop}
\begin{proof}
Following from the assumption in the proposition, 
\begin{align}
    \bY_\bA & = \bA\bY + \textbf{b} = \bA(\bxi + \bomega \bZ_0) + \textbf{b} = \bxi_\bA + \bA\bomega \bZ_0  = \bxi_\bA + (\bU_1^\bA|\bU_0 + \btau > \0), \label{eq8} 
\end{align}
where $\bU_1^\bA = \bA\bomega\bU_1$. From \eqref{eq8}, the random vector $$
\begin{pmatrix}
    \bU_0 \\
    \bU_1^\bA
\end{pmatrix} \sim {\cal T}_{m+n}\left(\begin{pmatrix}
    \0 \\
    \0
\end{pmatrix}, \begin{pmatrix}
    \Bar{\bGamma} & \bDelta^\top\bomega\bA^\top\\
    \bA\bomega\bDelta & \bA\bOmega\bA^\top
\end{pmatrix}, \nu \right) = {\cal T}_{m+n}\left(\begin{pmatrix}
    \0 \\
    \0
\end{pmatrix}, \begin{pmatrix}
    \Bar{\bGamma} & \bDelta_\bA^\top \bomega_\bA\\
    \bomega_\bA\bDelta_\bA & \bOmega_\bA
\end{pmatrix}, \nu \right).
$$
Again, by the stochastic representation \eqref{eq all}, $\bY_\bA$ follows a SUT distribution with transformed parameters as specified above. 
\end{proof}
Similar to the SUN distribution, the latent parameters $\btau$ and $\bar \bGamma$ of a SUT distribution remain unchanged under linear transformations. Moreover, the degrees-of-freedom parameter remains unchanged too. With the general linear transformation property, we can formulate a series of propositions involving additivity, marginal distributions, and canonical form.

\subsubsection{Marginal Distribution} \label{5.1}

In this section, we show that the marginals of a SUT distribution still follow a SUT distribution. For this purpose, we assume that the random vector $\bY \sim {\cal SUT}_{d,m}(\bxi,\bOmega,\bDelta,\btau,\Bar{\bGamma}, \nu)$ can be partitioned as
\begin{align} \label{partition}
    \bY = \begin{pmatrix}
        \bY_1 \\
        \bY_2
    \end{pmatrix} \sim {\cal SUT}_{d_1 + d_2,m}\left(\begin{pmatrix}
        \bxi_1 \\
        \bxi_2
    \end{pmatrix},\begin{pmatrix}
        \bOmega_{11} & \bOmega_{12}\\
        \bOmega_{21} & \bOmega_{22}
    \end{pmatrix}, \begin{pmatrix}
        \bDelta_1 \\
        \bDelta_2
    \end{pmatrix},\btau,\Bar{\bGamma}, \nu \right),
\end{align}
where $\bY_i \in \R^{d_i}$, $\bxi_i \in \R^{d_i}$, $\bOmega_{ij} \in \R^{d_i \times d_j}$, $\bomega_i = \diag(\bOmega_{ii})^{1/2}$, $\bar\bOmega_{ij} = \bomega_i^{-1} \bOmega_{ij} \bomega_j^{-1}$ and $\bDelta_i \in \R^{d_i \times m}$, for $i, j \in \{1, 2\}$, such that $d_1 +\,d_2 = d$.
\begin{prop}\label{propMarginals}
    Let $\bY \sim {\cal SUT}_{d,m}(\bxi,\bOmega,\bDelta,\btau,\Bar{\bGamma}, \nu)$. Then, the marginal random vector $\bY_i \sim {\cal SUT}_{d_i,m}(\bxi_i,\bOmega_{ii},\bDelta_i,\btau,\Bar{\bGamma}, \nu)$, for $i \in \{1, 2\}.$
\end{prop}
\begin{proof}
    Consider the marginal random vector $\bY_1 = \bA_1\bY$, where $\bA_1 = (
    \bI_{d_1},\0)$ denotes the projection matrix on the first $d_1$ coordinates. After applying the properties of linear transformations derived in Proposition  \ref{proplinear}, it follows that $\bY_1 \sim {\cal SUT}_{d_1,m}(\bxi_1,\bOmega_{11},\bDelta_1,\btau,\Bar{\bGamma}, \nu)$ because $\bA_1\bxi = \bxi_1$, $\bA_1\bOmega\bA_1^\top = \bOmega_{11}$, and $\bomega_{\bA_1}^{-1}\bA_1\bomega\bDelta = \bomega_{1}^{-1}\bomega_{1}\bDelta_1 = \bDelta_1$, where $\Bar{\bOmega}_{11} = \bomega_{1}^{-1}\bOmega_{1}\bomega_{1}^{-1}$. Analogously, one obtains the distribution of $\bY_2 = \bA_2 \bY \sim {\cal SUT}_{d_2,m}(\bxi_2,\bOmega_{22},\bDelta_2,\btau,\bGamma,\nu)$, with $\bA_2 = ( \0, \bI_{d_2}).$
\end{proof}
Note from Proposition \ref{cov} that each marginal can be represented stochastically as 
$$\bY_i=\bxi_i+\bomega_i\bZ_i,\quad
\bZ_i=\bDelta_i\bar\bGamma^{-1}\bU_*+\sqrt{\frac{\nu + Q_{\bU_*}}{\nu + m}} \bW_{*i},\quad i=1,2,$$
where $ \bW_{*i}=\bA_i\bW_*\sim{\cal T}_{d_i}(\0,\bar\bOmega_{ii}-\bDelta_i\bar\bGamma^{-1}{\bDelta_i}^\top,\nu)$ is independent of $\bU_*$. From this result it is clear that if $\bDelta_i=\0$, then $\bY_i$ will have a symmetric distribution, which becomes a multivariate~$t$ when $\btau=\0$. Moreover, provided that the required moments exist, we have:
\begin{align*}
\mbox{Cov}(\bY_1,\bY_2) & =\bomega_1\mbox{Cov}(\bZ_1,\bZ_2)\bomega_2\\
& = \bomega_1\mbox{Cov}\left(\bDelta_1\bar\bGamma^{-1}\bU_*+\sqrt{\frac{\nu + Q_{\bU_*}}{\nu + m}} \bW_{*1},\bDelta_2\bar\bGamma^{-1}\bU_*+\sqrt{\frac{\nu + Q_{\bU_*}}{\nu + m}} \bW_{*2}\right)\bomega_2\\
& = \bomega_1\bDelta_1\bar\bGamma^{-1}\mbox{Var}(\bU_*)\bar\bGamma^{-1}\bDelta_2^\top\bomega_2
+ \eta(Q_{\bU_*})\bomega_1\mbox{Cov}(\bW_{*1},\bW_{*2})\bomega_2,
\end{align*}
where $\eta(Q_{\bU_*})$ is defined in (\ref{etaQ}). We can then conclude that $\mbox{Cov}(\bY_1,\bY_2)=\0$ if any of the following conditions are met: (i) $\bDelta_1=\0$ and $\mbox{Cov}(\bW_{*1},\bW_{*2})=\0$; or (ii) $\bDelta_2=\0$ and $\mbox{Cov}(\bW_{*1},\bW_{*2})=\0$. 
Since $\mbox{Cov}(\bW_{*1},\bW_{*2})=\frac{\nu}{\nu-2}\left(\bar\bOmega_{12}-\bDelta_1\bar\bGamma^{-1}{\bDelta_2}^\top\right)$, it follows that $\bY_1$ and $\bY_2$ are uncorrelated under the condition that  $\bar\bOmega_{12}=\0$ and either $\bDelta_1=\0$ or $\bDelta_2=\0$.


\subsubsection{Additivity}
A specific case of the linear transformation is the additivity of the marginals $\bY = (\bY_1^\top,\bY_2^\top)^\top$, where $\bY_1 \in \R^d$ and $\bY_2 \in \R^d$. We have the following proposition. 
\begin{prop}
Let $\bY_1 \in \R^{d}$ and $\bY_2 \in \R^{d}$ be two random vectors with joint distribution:
\begin{equation} \label{4.2d}
    \begin{pmatrix}
        \bY_1 \\
        \bY_2
    \end{pmatrix} \sim {\cal SUT}_{d + d, m}\left(
    \begin{pmatrix}
        \bxi_1 \\
        \bxi_2
    \end{pmatrix},
    \begin{pmatrix}
        \bOmega_{11} & \bOmega_{12} \\
        \bOmega_{21} & \bOmega_{22}
    \end{pmatrix}, \begin{pmatrix}
        \bDelta_{1} \\ 
        \bDelta_{2}    
    \end{pmatrix},\btau,
    \Bar{\bGamma}, \nu
    \right).
\end{equation}
Then $\bY_1 + \bY_2$ is ${\cal SUT}_{d, m}(\bxi_+, \bOmega_+, \bDelta_+,\btau,\Bar{\bGamma},\nu)$ with $\bxi_+ = \bxi_1 + \bxi_2$, $\bOmega_+ = \bOmega_{11} + \bOmega_{22}+ \bOmega_{12}+ \bOmega_{21}$, $ \bomega_+=\mbox{diag}(\bOmega_+)$, $\bDelta_+ =\bomega_+^{-1}\left(\bomega_1\bDelta_{1}+\bomega_2\bDelta_{2}\right)$
where $\bOmega_+ = \bomega_+ \Bar{\bOmega}_+ \bomega_+$,  $\bOmega_{11} = \bomega_1 \Bar{\bOmega}_{11} \bomega_1$, and $\bOmega_{22} = \bomega_2 \Bar{\bOmega}_{22} \bomega_2$. 
\end{prop}
\begin{proof}
The derivation for the additivity is through the properties of linear transformations demonstrated in Proposition \ref{proplinear}. If we let $\bA = (\bI_d,\bI_d)$ and $\textbf{b} = \0$, then
\begin{align*}
    \bxi_+ & = \bA \begin{pmatrix}
        \bxi_1 \\
        \bxi_2
    \end{pmatrix} = \bxi_1 + \bxi_2, \quad
    \bOmega_+  = \bA \begin{pmatrix}
        \bOmega_{11} & \bOmega_{12}\\
        \bOmega_{21} & \bOmega_{22}
        \end{pmatrix}\bA^\top = \bOmega_{11} + \bOmega_{22}+ \bOmega_{12}+ \bOmega_{21}, \\
        \bDelta_+ & = \bomega_+^{-1}\bA \begin{pmatrix}
            \bomega_1 & \0 \\
            \0 & \bomega_2
        \end{pmatrix} \begin{pmatrix}
            \bDelta_{1}\\
            \bDelta_{2}
        \end{pmatrix} = \bomega_+^{-1}\left(\bomega_1\bDelta_{1}+\bomega_2\bDelta_{2} \right).
\end{align*}
\end{proof}
 Therefore, the sum of two SUT random vectors still follows a SUT distribution given that their joint distribution has the structure indicated in \eqref{4.2d}. However, unlike the SUN distribution, in which the sum of any two independent SUN random vectors are closed under the same distribution, the assumption that $(\bY_1^\top,\bY_2^\top)^\top$ follows a SUT distribution is needed \citep{arellano2010multivariate}. Thus, only the sum of the marginals of a SUT random vector is closed under the same distribution. Some conditions to have a null correlation between $\bY_1$ and $\bY_2$ are given in the previous subsection. For instance, this fact holds if $\bOmega_{12}=\0$ and $\bDelta_1=\0$ or
 $\bDelta_2=\0$.

However, uncorrelation of the marginals is not guaranteed for finite values of  $\nu$ if $\bDelta_1 \neq \0$ and $\bDelta_2 \neq \0$, because the truncation operation introduces correlation between the corresponding marginals through $\bU_*$. In fact, consider $(\bY_1^\top, \bY_2^\top)^\top$ partitioned as in \eqref{4.2d} and, additionally, assume that  $\bOmega_{12}=\0$, $\bDelta_1=(\begin{matrix}\bDelta_{11}&\0\end{matrix})$, $\bDelta_2=(\begin{matrix}\0 &\bDelta_{22}\end{matrix})$ and 
$\bar\bGamma=\mbox{diag}(\bar\bGamma_{11},\bar\bGamma_{22})$, with respective latent dimensions $m_1$ and $m_2$ $(m=m_1+m_2)$, and $\btau = \0$. So, by Proposition \ref{p3}, we have $\bY_i = \bxi_i + V^{-1/2} \bZ_i$, where $V \sim {\cal G}amma(\nu/2, \nu/2)$ and $\bZ_i \sim \SUN_{d, m_i}(\0, \bOmega_{ii}, \bDelta_{ii}, \0, \bar\bGamma_{ii})$, for $i \in \{1,2\}$. Here $V$ is independent with respect to each $\bZ_i$ and, by Appendix B in \cite{arellano2006unification}, $\bZ_1$ and $\bZ_2$ are independent as well. Thus, we end up with
\begin{align*}
    \mbox{Cov}(\bY_1,\bY_2) = (M_2 - M_1^2) \, \E(\bZ_1) \E(\bZ_2),
\end{align*}
where $M_1 = \E(V^{-1/2}) = \sqrt{(\nu/2)}\, \Gamma\{(\nu - 1)/2\}/\Gamma(\nu/2)$ and $M_2 = \E(V^{-1}) = \nu/(\nu -2)$. By equation (8) in \cite{arellano2022some}, $\E(\bZ_i)$ only vanishes if $\bDelta_{ii} = \0$. Moreover, $\lim_{\nu \to \infty} (M_2 - M_1^2) = 0$, so that $\bY_1$ and $\bY_2$ are uncorrelated only in the limiting case, i.e., when $\nu \to \infty$, which brings us back to the SUN case.

\subsubsection{Canonical Form}
An interesting result based on Proposition \ref{cov} is the formulation of the canonical form \citep{Capi2020} of a SUT random vector. The canonical form moves all of the skewness of a SUT random vector toward the first component, leaving the remaining components symmetric. 
\begin{prop}\label{prop:canon}
    Let $\bY \sim {\cal SUT}_{d,m}(\bxi,\bOmega,\bDelta,\btau,\Bar{\bGamma},\nu)$ be partitioned as $(\text{Y}_1,\bY_2^\top)^\top$, where $Y_1 \in \R$ and $\bY_2 \in \R^{d-1}$. Then, the representation in Proposition \ref{cov} with $\bDelta = (\bDelta_1^\top,\bDelta_2^\top)^\top$ and $\bDelta_2 = \0$ results in the canonical form, where $\bDelta_1 \in \R^{m}$ and $\bDelta_2 \in \R^{(d-1) \times m}$. 
\end{prop}
\begin{proof}
    Using the results from Proposition \ref{cov}, we have $\bY  = \bxi + \bomega \left (\bB_0 \bU_* + \sqrt{\frac{\nu + Q_{\bU_*}}{\nu + m}} \bW_* \right)$ and 
\begin{align}
    \begin{pmatrix}
        Y_1 \\
        \bY_2
    \end{pmatrix} & = \bxi + \bomega \left\{ \begin{pmatrix}
        \bDelta_1\Bar{\bGamma}^{-1} \\
        \0
    \end{pmatrix}  \bU_* + \sqrt{\frac{\nu + Q_{\bU_*}}{\nu + m}} \bW_* \right \}. \label{canon}
\end{align}
In (\ref{canon}), the truncated $t$ random vector is only distributed to the first component $Y_1$ and the quantity is $\bDelta\Bar{\bGamma}^{-1}\bU_*$. The $(d-1) \times m$ zero matrix $\bDelta_2$ annihilates the skewness in $\bY_2$. Therefore, $Y_1$ is a univariate SUT random variable and $\bY_2 \in \R^{d - 1}$ is a symmetric multivariate random vector of the form $\sqrt{\frac{\nu + Q_{\bU_*}}{\nu + m}} \bW_{*2}$, where $\bW_* = (W_{*1},\bW_{*2}^\top)^\top$. As mentioned in the previous subsection, we note that $\sqrt{\frac{\nu + Q_{\bU_*}}{\nu + m}} \bW_{*2}$ will be a multivariate $t$ when $\btau = \0$. 
\end{proof}
With Propositions \ref{proplinear} and \ref{prop:canon}, it is possible to construct the canonical (linear) transformation matrix $\bA = \textbf{C} = \mbox{diag}(C_1,\textbf{C}_2)$ for the regular SUT random vector $\bY$, where $C_1 \in \R$ and $\textbf{C}_2 \in \R^{(d - 1) \times (d - 1)}$ such that $\textbf{C}_2 \bDelta_{2} = \0$, so that $\textbf{C}\bY$ has the canonical form. Note that the canonical transformation matrix $\textbf{C}$ requires the column space of $\bDelta_2$ to be a subspace of the null space of $\textbf{C}_2$. Hence, the canonical transformation does not always exist. This result is different from the SN family, for which the canonical form always exists \citep{Capi2020}, which is equivalent to $\SUN_{d,1}$ or ${\cal SUT}_{d,1}$. The assured existence of the canonical transformation for the SN family is due to the fact that it is always possible to construct a transformation matrix with one-dimensional null space. In the SUT case ($m \geq 2$), the matrix $\bDelta_2 \in \R^{(d-1)\times m}$. Hence, rank$(\bDelta_2) \leq$ min($d-1,m$). The null space for $\textbf{C}_2 \in \R^{(d-1) \times (d-1)}$ is at most $(d - 2)$-dimensional. Hence, we need $m \leq d-2$ in order for the canonical transformation to exist. The canonical form and transformation for the SUN are obtained when $\nu\rightarrow \infty$.

\subsection{Quadratic Forms} \label{4.3}

\cite{arellano2010invariance} have systematically explored the conditions for which the invariance property of the quadratic form holds for selection distributions. The main motivation was to find the condition that makes the two random vectors in the selection representation uncorrelated and therefore, the conditional cumulative distribution evaluated at the center is identical with the the unconditional cumulative distribution of the component under the same circumstances. Following this approach, in this section we compute the quadratic form of the SUT random vector and explore the required conditions that establish the invariance property for the SUT.  

According to \cite{ABG2006}, one of the main properties of selection distributions is: if $\bZ=(\bU_1|\bU_0\in C)$ then $ g(\bZ)=(g(\bU_1)|\bU_0\in C)$ for every measurable function $g: \R^d\to\R^p$. It follows that, for example, if the distribution of $\bU_1$ is closed under linear transformations then the distribution of $\bZ$ is also closed. Note that if $g(\bU_1)$ and $\bU_0$ are uncorrelated (which can happen when $(\bU_0,\bU_1)$ is symmetric about the origin and $g(\cdot)$ is an even function), then we may have that $\Prob(\bU_0\in C| g(\bU_1)=\bx)=\Prob(\bU_0\in C)$.

With the aforementioned framework and property, we provide detailed steps to calculate the pdf of quadratic forms in SUT random vectors. 
\begin{prop} \label{quad}
    Let $ \bY \sim {\cal SUT}_{d,m}(\bxi,\bOmega,\bDelta,\btau,\Bar{\bGamma},\nu)$. The quadratic form $Q_\bY = (\bY-\bxi)^\top\bOmega^{-1}(\bY - \bxi)$ has pdf
    $$
    f_{Q_{\bY}}(v)= f_{Q_{\bU_1}}(v)  \frac{\Prob(\bU_0^*<\bLambda \bU_1 + \btau |Q_{\bU_1} = v)}{\Prob(\bU_0^* < \bLambda \bU_1 + \btau)},
    $$    
where $Q_{\bU_1} = \bU_1^\top \Bar{\bOmega}^{-1}\bU_1$ with pdf $f_{Q_{\bU_1}}(v)$ and $\bLambda = \bDelta^\top \Bar{\bOmega}^{-1}$.
\end{prop}
\begin{proof}
Consider the stochastic representation $\bY = \bxi + \bomega\bZ_0$, where $\bZ_0=(\bU_1|\bU_0 + \btau > \0)$ and note that $Q_\bY = Q_{\bZ_0} = \bZ_0^\top\Bar{\bOmega}^{-1}\bZ_0$. Then:
\begin{align*}
    Q_{\bZ_0} 
     = \left(\bU_1^\top\Bar{\bOmega}^{-1}\bU_1|\bU_0 + \btau > \0 \right) 
    =\left(\bU_1^\top\Bar{\bOmega}^{-1}\bU_1|\bU_0^* < \bLambda\bU_1 + \btau \right) 
     = \left(Q_{\bU_1}|\bU_0^* < \bLambda\bU_1 + \btau \right),
\end{align*}
where $\bU_0^*=\bLambda\bU_1-\bU_0$, which is uncorrelated with $\bU_1$ (see (\ref{eq3})).
Hence, the pdf of the quadratic form has the form stated.
\end{proof}
To compute the conditional probability $\Prob(\bU_0^* < \bLambda \bU_1 + \btau |Q_{\bU_1} = v)$, we need two well-known results from \cite{fang2018symmetric}.
First, if $\bX_1$ and $\bX_2$ are jointly elliptical and uncorrelated, that is $
    (\bX_1^\top,\bX_2^\top)^\top
\sim {\cal EC}_{d_1+d_2}(\bmu,\bSigma,h)$ with location parameter $\bmu=(\bmu_1^\top,\bmu_2^\top)^\top$, dispersion matrix $\bSigma = \mbox{diag}(\bSigma_{11},\bSigma_{22})$,  and density generator function $h$,
then $(\bX_1|\bX_2) = (\bX_1|Q_{\bX_2})$ and $(\bX_2|\bX_1) = (\bX_2|Q_{\bX_1})$, where $Q_{\bX_1} = \bX_1^\top\bSigma_{11}^{-1}\bX_1$ and $Q_{\bX_2} = \bX_2^\top \bSigma_{22}^{-1}\bX_2$. 
Second, it is well known that $\bW_{1} = Q_{\bX_1}^{-1/2}\bSigma_{11}^{-1/2}\bX_1$ is a uniform random vector on the unit sphere of $\R^{d_1}$  (denoted usually by $\bU^{(d_1)}$) and that it is independent of $Q_{\bX_1}$ and of $Q_{\bX_1}^{-1/2}\bX_2$. Applying these two results, we have:
\begin{align*}
    \Prob(\bU_0^* < \bLambda \bU_1 + \btau | Q_{\bU_1} = v) & = \mathbb{E}_{\bW_1}\{\Prob(\textbf{T}_* < \Bar{\bLambda}\bW_1 + \btau_v| \bW_1, Q_{\bU_1} = v)\} 
\end{align*}
where $\textbf{T}_* = Q_{\bU_1}^{-1/2}\bU_0^*$, $\Bar{\bLambda} = \bLambda \Bar{\bOmega}^{1/2}$, $\btau_v = v^{-1/2}\btau$, $\mathbf{\Upsilon} = \Bar{\bGamma} - \bDelta^\top \Bar{\bOmega}^{-1}\bDelta$, and $\bW_1 = Q_{\bU_1}^{-1/2}\Bar{\bOmega}^{-1/2}\bU_1$. Note that, by the first result, $(\bU_0^* | Q_{\bU_1} = v)  \sim {\cal T}_{m}(\0,\alpha_{\nu,v}\mathbf{\Upsilon}, \nu + d)$. Therefore, $(\textbf{T}_*|Q_{\bU_1} = v) \sim {\cal T}_m(\0,v^{-1}\alpha_{\nu,v}\mathbf{\Upsilon}, \nu + d)$ and so $\Prob(\bU_0^* < \bLambda \bU_1 + \btau | Q_{\bU_1} = v)=\mathbb{E}_{\bW_1}\{T_m(\Bar{\bLambda}\bW_1 + \btau_v;v^{-1}\alpha_{\nu,v}\mathbf{\Upsilon},\nu + d)\}$.  Consequently,
\begin{align*}
    f_{Q_\bY}(v) = f_{Q_{\bU_1}}(v)\frac{\mathbb{E}_{\bW_1}\{T_m(\Bar{\bLambda}\bW_1 + \btau_v;v^{-1}\alpha_{\nu,v}\mathbf{\Upsilon},\nu + d)\}}{T_m(\btau;\Bar{\bGamma},\nu)}.
\end{align*}
In the pdf above, $Q_{\bU_1}$ does not have a closed form because, even though the square of a $t$ random variable follows an $F$-distribution, the sum of $F$ random variables does not yield a random variable with an explicit pdf. Moreover, because $\mathbb{E}_{\bW_1}\{T_m(\Bar{\bLambda}\bW_1 + \btau_v;v^{-1}\alpha_{\nu,v}\mathbf{\Upsilon},\nu + d)\}$ and $T_m(\btau;\Bar{\bGamma},\nu)$ must be computed numerically, the computation of $f_{Q_\bY}(v)$ has to rely on numerical methods. 

Another interesting point is that if $\Bar{\bLambda} = \0$ and $\btau_v = \0$, hence $\bDelta = \0$ and $\btau = \0$, then $\mathbb{E}_{\bW_1}\{T_m(\Bar{\bLambda}\bW_1 + \btau_v;v^{-1}\alpha_{\nu,v}\mathbf{\Upsilon},\nu + d)\} = T_m(\0;v^{-1}\alpha_{\nu,v}\Bar{\bGamma},\nu +d ) = T_m(\0;\alpha_{\nu,v}\Bar{\bGamma},\nu +d ) = \Prob(\bU_0^* < \0|\bU_1)$. Furthermore, according to \cite{fang2018symmetric}, if the marginals of the joint elliptical distribution are uncorrelated then $\Prob(\bU_0^* < \0|\bU_1)  = \Prob(\bU_0^* < \0) = \Phi_m(\0;\0,\Bar{\bGamma})$. Therefore, $f_{Q_\bY}(v) = f_{Q_{\bU_1}}(v)$. However, when $\bDelta = \0$ and $\btau=\0$, the random vector $\bY$ has a multivariate $t$ distribution and, therefore, is no longer skewed. 

This result can also be directly obtained from the $f_{Q_\bY}(v)$ in Proposition \ref{quad}. In particular, $\Prob(\bU_0^* < \bLambda\bU_1 + \btau|Q_{\bU_1} = v) = \Prob(\bU_0^* < \bLambda \bU_1 + \btau)$ if $\text{Cov}(\bU_1, \bU_0^* - \bLambda\bU_1) = \0$ and $\btau = \0$.  We can observe that 
$\text{Cov}(\bU_1, \bU_0^* - \bLambda\bU_1) = \text{Cov}(\bU_1, \bU_0^* ) - \text{Cov}(\bU_1, \bLambda\bU_1) = -\bDelta = \0$.

In addition, when $\btau=\0$ and $m\ge 2$, another way to study the distribution of $Q_\bZ=\bZ^\top\bar\bOmega^{-1}\bZ$, where $\bZ=(\bU_1|\bU_0>\0)$, is to use the representation $Q_\bZ=V^{-1}Q_{\bZ_0}$, where $V$ is independent of $Q_{\bZ_0}=\bZ_0^\top\bar\bOmega^{-1}\bZ_0$, with $\bZ_0=(\bW_1|\bW_0>\0)$ and $(\bW_0,\bW_1)$ being the respective (centered) normal variables. Thus, since $\bar\bW_1=\bW_1-\bLambda\bW_0$, where $\bLambda= \bDelta\bar\bGamma^{-1}$, is independent of $\bW_0$, we have $Q_{\bZ_0}=(\bar\bW_1+\bLambda|\bW_0|)^\top\bar\bOmega^{-1}(\bar\bW_1+\bLambda|\bW_0|)$, which has $\chi_{d}^2$ distribution (and so $Q_\bZ$ has a Fisher type of distribution for $m\ge2$) iff $\bLambda= \bDelta\bar\bGamma^{-1}=\0$. When $m=1$, we know that $Q_{Z_0}\sim\chi_1^2$ \citep{AC2014}. 

\vspace{-.1cm}

\section{Conditional Distributions} \label{s5}

In this section, we show that the conditionals of a SUT distribution still follow a SUT distribution, and that a form of conditioning allows to change the latent dimension.

\subsection{Conditional Distribution} \label{5.2 con}

\begin{prop}\label{propConditionals}
    Let $\bY=(\bY_1^\top,\bY_2^\top)^\top \sim \SUT_{d,m}(\bxi, \bOmega, \bDelta, \btau, \bar\bGamma,\nu)$ be partitioned as in \eqref{partition}. Then:
\begin{align}
    (\bY_2 | \bY_1 = \by_1) \sim \SUT_{d_2, m}(\bxi_{2 \cdot 1}, \alpha_{\nu,Q_{\by_1}} \bOmega_{2 \cdot 1}, \bDelta_{2 \cdot 1}, \alpha_{\nu,Q_{\by_1}}^{-1/2} \btau_{2 \cdot 1}, \bar\bGamma_{2 \cdot 1}, \nu + d_1), \quad \by_1 \in \R^{d_1},
\end{align}
where $\bxi_{2 \cdot 1} = \bxi_2 + \bOmega_{21} \bOmega_{11}^{-1} (\by_1 - \bxi_1)$, $\alpha_{\nu,Q_{\by_1}} = \{\nu + Q_{\by_1}\}/\{\nu + d_1\}$, $Q_{\by_1} = (\by_1 - \bxi_1)^\top \bOmega_{11}^{-1} (\by_1 - \bxi_1)$, $\bOmega_{2 \cdot 1} = \bOmega_{22} - \bOmega_{21} \bOmega_{11}^{-1} \bOmega_{12}$, $\bDelta_{2 \cdot 1} = \bomega_{2 \cdot 1}^{-1} (\bomega_2 \bDelta_2 - \bOmega_{21} \bOmega_{11}^{-1} \bomega_1 \bDelta_1) \bgamma_{2 \cdot 1}^{-1}$, $\bomega_{2 \cdot 1} = \diag(\bOmega_{2 \cdot 1})^{1/2}$,  $\btau_{2 \cdot 1} = \bgamma_{2 \cdot 1}^{-1}\{\btau +\bDelta_1^\top \bar\bOmega_{11}^{-1} \bomega_1^{-1} (\by_1 - \bxi_1)\}$, $\bar\bGamma_{2 \cdot 1} = \bgamma_{2 \cdot 1}^{-1} \bGamma_{2 \cdot 1} \bgamma_{2 \cdot 1}^{-1}$, $\bGamma_{2 \cdot 1} = \bar\bGamma - \bDelta_1^\top \bar\bOmega_{11}^{-1} \bDelta_1$ and $\bgamma_{2 \cdot 1} = \diag(\bGamma_{2 \cdot 1})^{1/2}$.
\end{prop}

\begin{proof}
    The proof follows the same reasoning as the one given for Proposition 3.2 in \cite{arellano2010multivariate} in the particular case where the Student's $t$ density generator function is considered  and the parameterization in Definition \ref{defi1} is used. 
\end{proof}
It is worth noting that the degrees-of-freedom is increased after conditioning by the dimension of the conditioning vector, hence making the resulting SUT distribution closer to the SUN.

\subsection{Changing Latent Dimensions} \label{latentdim} 

We investigate ways to change the dimension of the latent variables.
\begin{prop}
Let $\bY = (\bY_1^\top, \bY_2^\top)^\top \sim \SUT_{d,m}(\bxi, \bOmega, \bDelta, \btau, \bar\bGamma, \nu)$ with parameters partitioned as in \eqref{partition}. Then,
\begin{align}\label{condtrunc1}
    (\bY_2 | \bY_1 > \0) \sim \SUT_{d_2, d_1 + m}(\bxi_2, \bOmega_{22}, \bDelta_{2 \neg 1}, \btau_{2 \neg 1}, \bar{\bGamma}_{2 \neg 1}, \nu),   
\end{align}
where 
\begin{align}
    \bDelta_{2 \neg 1} = 
    \begin{pmatrix}
       \bDelta_2 & \bar\bOmega_{21}
    \end{pmatrix}, \quad
    \bar{\bGamma}_{2 \neg 1} = 
    \begin{pmatrix}
        \bar\bGamma & \bDelta_1^\top \\
        \bDelta_1 & \bar\bOmega_{11}
    \end{pmatrix}, \quad
    \btau_{2 \neg 1} = 
    \begin{pmatrix}
        \btau\\
        \bomega_1^{-1} \bxi_1
    \end{pmatrix}.
\end{align}

\end{prop}

\begin{proof}
Firstly, note that, from Proposition \ref{propMarginals} it follows that $\bY_2 \sim \SUT_{d_2, m}(\bxi_2, \bOmega_{22}, \bDelta_2, \btau, \bar\bGamma, \nu)$ and $-\bY_1 \sim \SUT_{d_1,m}(-\bxi_1, \bOmega_{11}, -\bDelta_1, \btau, \bar\bGamma, \nu)$. Secondly, from Proposition \ref{propConditionals} (exchanging the subscripts 1 and 2) we obtain that: 
$$(-\bY_1 | \bY_2 = \by_2) \sim \SUT_{d_1, m}(-\bxi_{1 \cdot 2}, \alpha_{\nu, Q_{\by_2}} \bOmega_{1 \cdot 2}, -\bDelta_{1 \cdot 2}, \alpha_{\nu, Q_{\by_2}}^{-1/2} \btau_{1 \cdot 2}, \bar\bGamma_{1 \cdot 2}, \nu + d_2).$$ 
Hence:
\begin{align*}
    &f_{\bY_2}(\by_2) = t_{d_2}(\by_2; \bxi_2, \bOmega_{22}, \nu) \frac{T_m[\alpha_{\nu, Q_{\by_2}}^{-1/2} \{\btau + \bDelta_2^\top \bar\bOmega_{22}^{-1} \bomega_2^{-1} (\by_2 - \bxi_2)\}; \bar\bGamma - \bDelta_2^\top \bar\bOmega_{22}^{-1} \bDelta_2, \nu + d_2]}{T_m(\btau; \bar\bGamma, \nu)},\\
    &\Prob(-\bY_1 \leq \0 | \bY_2 = \by_2) = \frac{T_{d_1 + m}\left\{ \begin{pmatrix}
        \alpha_{\nu, Q_{\by_2}}^{-1/2} \btau_{1 \cdot 2} \\ \bxi_{1 \cdot 2}
    \end{pmatrix}; \begin{pmatrix}
        \bar\bGamma_{1 \cdot 2} & \bDelta_{1 \cdot 2}^\top \alpha_{\nu, Q_{\by_2}}^{1/2} \bomega_{1 \cdot 2}\\
        \bomega_{1 \cdot 2} \alpha_{\nu, Q_{\by_2}}^{1/2} \bDelta_{1 \cdot 2} & \alpha_{\nu, Q_{\by_2}} \bOmega_{1 \cdot 2}
    \end{pmatrix}, \nu + d_2 \right\}}{T_m(\alpha_{\nu, Q_{\by_2}}^{-1/2} \btau_{1 \cdot 2}; \bar\bGamma_{1 \cdot 2}, \nu + d_2)},\\
    &\Prob(-\bY_1 \leq \0) = \frac{T_{d_1 + m} \left\{\begin{pmatrix}
        \btau \\ \bxi_1
    \end{pmatrix}; \begin{pmatrix}
        \bar\bGamma & \bDelta_1^\top \bomega_1\\
        \bomega_1 \bDelta_1 & \bOmega_{11}
    \end{pmatrix}, \nu\right\}}{T_m(\btau; \bar\bGamma, \nu)}.
\end{align*}
Then, after some simplifications and a few algebraic manipulations, the pdf of $(\bY_2 | \bY_1 > \0)$ can be computed at $\by_2 \in \R^{d_2}$ as follows:
\begin{align*}
    &f_{\bY_2 | \bY_1 > \0}(\by_2) =  f_{\bY_2}(\by_2) \, \frac{\Prob(-\bY_1 \leq \0 | \bY_2 = \by_2)}{\Prob(-\bY_1 \leq \0)} \hspace{9cm} \\  
    &= t_{d_2}(\by_2; \bxi_2, \bOmega_{22}, \nu) \frac{T_{d_1 + m}\left\{ \begin{pmatrix}
        \alpha_{\nu, Q_{\by_2}}^{-1/2} \btau_{1 \cdot 2} \\ \bxi_{1 \cdot 2}
    \end{pmatrix}; \begin{pmatrix}
        \bar\bGamma_{1 \cdot 2} & \bDelta_{1 \cdot 2}^\top \alpha_{\nu, Q_{\by_2}}^{1/2} \bomega_{1 \cdot 2}\\
        \bomega_{1 \cdot 2} \alpha_{\nu, Q_{\by_2}}^{1/2} \bDelta_{1 \cdot 2} & \alpha_{\nu, Q_{\by_2}} \bOmega_{1 \cdot 2}
    \end{pmatrix},\nu + d_2 \right\}}{T_{d_1 + m} \left\{\begin{pmatrix}
        \btau \\ \bxi_1
    \end{pmatrix}; \begin{pmatrix}
        \bar\bGamma & \bDelta_1^\top \bomega_1\\
        \bomega_1 \bDelta_1 & \bOmega_{11}
    \end{pmatrix}, \nu\right\}}\\
    &= t_{d_2}(\by_2; \bxi_2, \bOmega_{22}, \nu) \frac{T_{d_1 + m}\left[\alpha_{\nu, Q_{\by_2}}^{-1/2} \{ \btau_{2 \neg 1} + \bDelta_{2 \neg 1}^\top \bar\bOmega_{22}^{-1} \bomega_2^{-1} (\by_2 - \bxi_2)\}; \bar\bGamma_{2 \neg 1} - \bDelta_{2 \neg 1}^\top \bar\bOmega_{22}^{-1} \bDelta_{2 \neg 1}, \nu + d_2\right] }{T_{d_1 + m} \left\{\btau_{2 \neg 1}; \bar\bGamma_{2 \neg 1}, \nu\right\}},
\end{align*} 
which, according to \eqref{pdfSUT}, is the pdf of a $\SUT_{d_2, d_1 + m}(\bxi_2, \bOmega_{22}, \bDelta_{2 \neg 1}, \btau_{2 \neg 1}, \bar{\bGamma}_{2 \neg 1}, \nu)$ distribution.
\end{proof}


Next, we explore the possibility of having redundant latent dimensions. 

\begin{lem} \label{uncorrelation}
    We have 
    \begin{equation}\label{splitT}
        T_{d + m}(\by_* - \bxi_*; \bOmega_*, \nu) = T_d(\by - \bxi; \bOmega, \nu)T_m(\0;\Bar{\bGamma}, \nu),    
    \end{equation}
    where $\by_* = (\0^\top,\by^\top)^\top$, $\bxi_* = (\0^\top, \bxi^\top)^\top$, and $\bOmega_* = \mbox{diag}(\Bar\bGamma, \bOmega)$.
\end{lem}
\begin{proof}
    On the one hand, if $\bY \sim {\cal SUT}_{d,m}(\bxi, \bOmega, \0, \0, \Bar{\bGamma}, \nu)$, then $\bY$ is elliptically distributed. So, it holds that $\Prob(\bY \leq \by) = T_d(\by - \bxi; \bOmega, \nu).$
    On the other hand, from \eqref{cdf}, we have $\Prob(\bY \leq \by) = T_{d + m}(\by_* - \bxi_*; \bOmega_*, \nu)/T_m(\0;\Bar{\bGamma}, \nu).$ Thus, \eqref{splitT} follows by equating the two previous identities.
\end{proof}
\begin{rem}
Equation \eqref{splitT} holds more generally. Indeed, let $(\bX_0,\bX_1)=R\bU^{(m+d)}=(R_0\bU^{(m)},R_1\bU^{(d)})$ be an $(m+d)$-dimensional spherical random vector, where $R=\sqrt{R_0^2+R_1^2}$ (the radial variable) and $\bU^{(m+d)}$ (the uniform vector on the unit sphere) are independent; also $(R_0,R_1)$, $\bU^{(m)}$ and $\bU^{(d)}$ are independent. Hence:
\begin{align*}\Prob(\bX_0\leq \0,\bX_1\leq \by)&=\Prob(R_0\bU^{(m)}\leq \0, R_1\bU^{(d)}\leq \by)=\Prob(\bU^{(m)}\leq \0, R_1\bU^{(d)}\leq \by)\\&
=\Prob(\bU^{(m)}\leq \0)\Prob( R_1\bU^{(d)}\leq \by)=\Prob(R_0\bU^{(m)}\leq \0)\Prob( R_1\bU^{(d)}\leq \by)\\&=\Prob(\bX_0\leq \0)\Prob( \bX_1\leq \by).
\end{align*}
It then follows for the uncorrelated elliptically contoured case defined by  $\bY_0=\bOmega_0^{1/2}\bX_0$ and $\bY_1=\bOmega_1^{1/2}\bX_1$, i.e.,  $(\bY_0^\top,\bY_1^\top)^\top\sim {\cal EC}_{m+d}(\0,\mbox{diag}(\bOmega_0,\bOmega_1),h^{(m+d)})$ where $h^{(m+d)}$ is the density generator, that also $\Prob(\bY_0\leq \0,\bY_1\leq \by)=\Prob(\bY_0\leq \0)\Prob( \bY_1\leq \by)$.
\end{rem}

The result in \eqref{splitT} can be used to show the following.
\begin{prop}\label{prop_lat_dim_reducer}
 If $\bY \sim {\cal SUT}_{d, m_1 + m_2}(\bxi, \bOmega, \bDelta, \btau, \Bar{\bGamma}, \nu)$ with $\bDelta = (\0, \bDelta_2)$, $\btau = (\0^\top, \btau_2^\top)^\top$, and $\Bar{\bGamma} = \diag(\Bar{\bGamma}_{11}, \Bar{\bGamma}_{22})$, then $\bY \sim {\cal SUT}_{d, m_2}(\bxi, \bOmega, \bDelta_2, \btau_2, \Bar{\bGamma}_{22},\nu)$.
\end{prop}

\begin{proof}
We have for the cdf of $\bY$:
\begin{align*}
F_{\bY}(\by) & = \frac{T_{d+m_1 + m_2}(\by^* - \bxi^*; \bOmega^*, \nu)}{T_{m_1+m_2}((\0^\top, \btau_2^\top)^\top;\Bar{\bGamma}, \nu)}, \quad \by^* = (\0^\top, \btau_2^\top, \by^\top)^\top,\quad \bxi^* = (\0^\top, \0^\top, \bxi^\top)^\top, \\
  &\hspace{65mm} \bOmega^* = \begin{pmatrix}
   \Bar{\bGamma}_{11} & \0 & \0 \\
   \0 & \Bar{\bGamma}_{22} & - \bDelta_2^\top \bomega \\
   \0 & - \bomega\bDelta_2 & \bOmega \\
\end{pmatrix}, \\
& = \frac{T_{m_1}(\0;\Bar{\bGamma}_{11}, \nu) \, T_{d + m_2}(\by^- - \bxi^-; \bOmega^-, \nu)}{T_{m_1}(\0; \Bar{\bGamma}_{11}, \nu)\, T_{m_2}(\btau_2; \Bar{\bGamma}_{22}, \nu)},  \quad \by^- = (\btau_2^\top, \by^\top)^\top,\quad \bxi^- =(\0^\top, \bxi^\top)^\top,\\
& \hspace{80mm} \bOmega^- = \begin{pmatrix}
    \Bar{\bGamma}_{22} & -\bDelta_2^\top \bomega \\
    -\bomega \bDelta_2 & \bOmega \\
\end{pmatrix}, \\
& = \frac{T_{d + m_2}(\by^- - \bxi^-; \bOmega^-, \nu)}{T_{m_2}(\btau_2; \Bar{\bGamma}_{22}, \nu)},
\end{align*}
which is the cdf of a ${\cal SUT}_{d, m_2}(\bxi, \bOmega, \bDelta_2, \btau_2, \Bar{\bGamma}_{22},\nu)$ distribution.
\end{proof}
Consequently, following from this property, if the random vector $\bY \sim {\cal SUT}_{d,m_1+ \dots+ m_n}(\bxi, \bOmega, \bDelta, \\ \btau, \Bar{\bGamma}, \nu)$, where $m = m_1 + \dots + m_n$, $\bDelta = (\bDelta_1,\dots,\bDelta_n)$, $\btau = (\btau_1^\top,\dots,\btau_n^\top)^\top$, and $\Bar{\bGamma} = \text{diag}(\Bar{\bGamma}_1, \dots, \Bar{\bGamma}_n)$, we can construct a latent dimension reduction matrix $\bR_i, i = 2, \dots,n$, for the dimension $m_i$ by solving the equation $\bR_i\bDelta_i = \0$, provided that $\btau_i = \0$. As a result, $\bR_i\bY \sim {\cal SUT}_{d,m - m_i}(\bR_i\bxi,\bR_i\bOmega\bR_i^\top,\bR_i\bDelta_{-i},\btau_{-i},\Bar{\bGamma}_{-i},\nu)$ with the negative index indicating the removal of the $i$-th element. Note that $i$ can be any number between 1 and $n$ because the SUT distribution is non-identifiable with respect to its latent variables \citep{nonke}. 

\section{Mardia's Measure of Multivariate Skewness and Kurtosis}\label{s7}

\subsection{Computation}
 
Mardia's measure of multivariate skewness and kurtosis \citep{Mardia1970} can also be computed exactly. Following from the previous setting, we consider $\bY \sim {\cal SUT}_{d,m}(\bxi, \bOmega, \bDelta, \btau, \Bar{\bGamma}, \nu)$ and denote $\text{Var}({\bY}) = \bSigma = \bL\bL^\top$ and $\mathbb{E}({\bY}) = \bmu = \bxi + \mathbb{E}(\bU_1^*|\bU_0 + \btau > \0) = \bxi + {\bmu}_0$, where $\bU_1^*=\bomega\bU_1$. 
We first need to standardize the random vector ${\bY}$ to compute the skewness and kurtosis measure. In particular, we let ${\bZ} = \bL^{-1}({\bY} - \bmu) \sim {\cal SUT}_{d,m}(-\bL^{-1}\tilde{\bmu}_0,\bOmega_\bL, \bDelta_\bL, \btau,\Bar{\bGamma}, \nu)$, where $\bOmega_\bL = \bL^{-1}\bOmega({\bL^{-1}})^\top$, $\bDelta_\bL = \bomega_\bL^{-1}\bL^{-1}\bomega\bDelta$ and $\bomega_\bL = \text{diag}(\bOmega_\bL)^{1/2}$. As a result, we have that $\mathbb{E({\bZ})} = \0$ and $\text{Var}({\bZ}) = \bI_d$. 

According to \cite{kollo2005estimation}, the Mardia measures of multivariate skewness and kurtosis of the standardized random vector $\bZ$ can then be computed using trace operation on the third and fourth moments:
\begin{align*}
    \beta_{1,d} & = \text{tr}\{\mu_3({\bZ})^\top\mu_3({\bZ})\} = \text{vec}\{\mu_3({\bZ})\}^\top\text{vec}\{\mu_3({\bZ})\}, \\
    \beta_{2,d} & = \text{tr}\{\mu_4({\bZ})\}.
\end{align*}

Here $\mu_3({\bZ})$ and $\mu_4({\bZ})$ can be computed using the convolution-based method described in Section \ref{con-based compute}. One point to notice is the displacement in the kurtosis measure when dealing with high dimensions. The non-shifted measure $\gamma_{2,d}$ can be adjusted by equation (2.9) in \cite{mardia1974applications}. Overall, $\gamma_{1,d}$ and $\gamma_{2,d}$ are invariant with respect to location and scale. Consequently, it is sufficient to assume that $\bY \sim {\cal SUT}_{d,m}(\0,\bar\bOmega,\bDelta,\btau,\bar\bGamma,\nu)$ as indicated in \cite{arellano2022some} for the computations.

\subsection{Visualization}

\begin{figure}[h!]
    \centering
    \includegraphics[width=0.8\textwidth,]{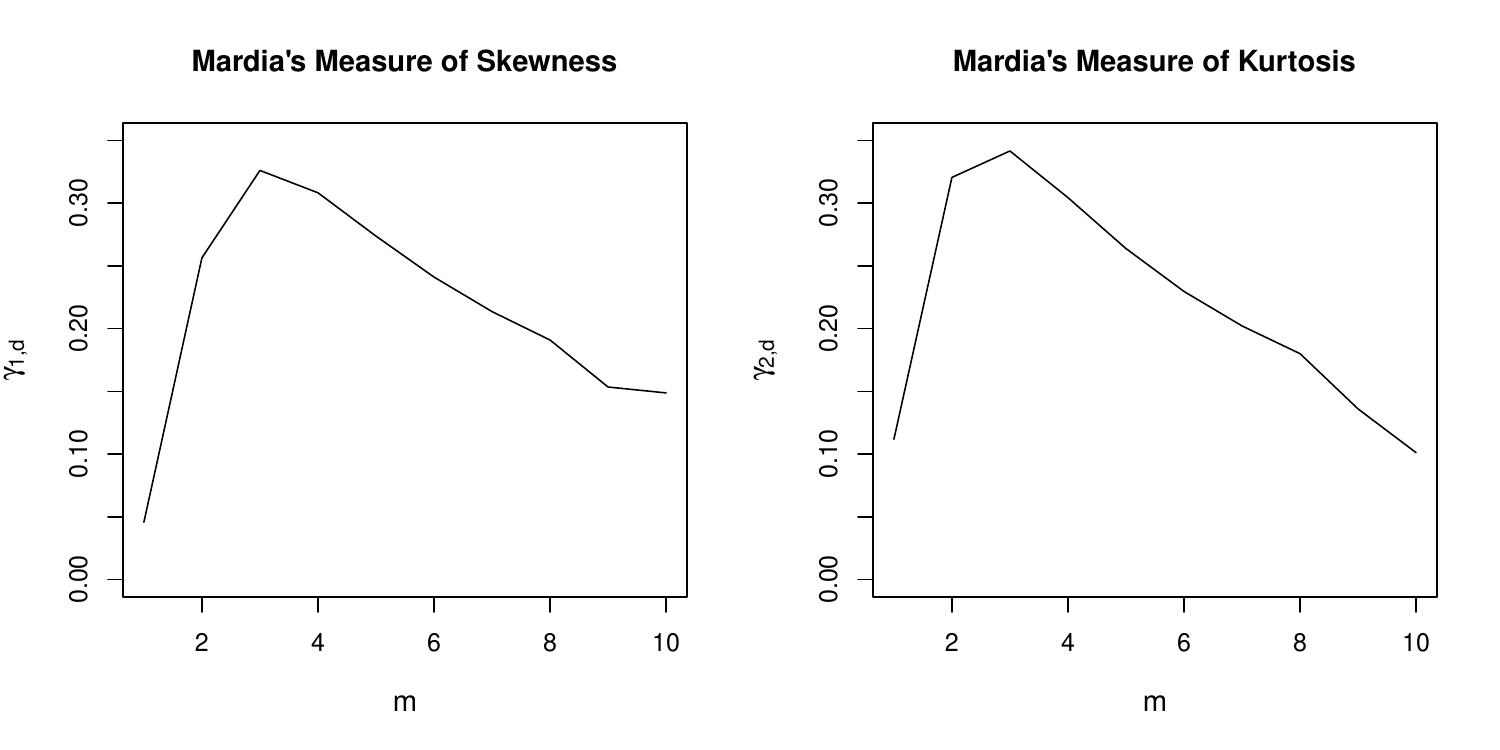}
    \caption{Mardia's measure of multivariate skewness $\gamma_{1,d}$ and kurtosis $\gamma_{2,d}$ against latent dimension $m$ with $(\bOmega,\bDelta,\bar\bGamma)$ specified to skew the distribution in the direction $(1,1)^\top$. Here $\bxi$ and $\btau$ are set as $\0$, and $\nu = 5$.}
    \label{mardia}
\end{figure}

To demonstrate the effectiveness of the two measures, we visualize $\beta_{1,d} = \gamma_{1,d}$ and $\gamma_{2,d}$ with increasing latent dimensions $m$ in Figure \ref{mardia}. We impose skewness to the distribution along the direction $(1,1)^\top$ to see the variations. 

Figure \ref{mardia} indicates that the two measures increase first and then decrease. The reason behind such a behavior is the asymptotic linear dependence in increasing latent dimensions. We articulate the rationales in the following proposition. 
\begin{prop} \label{lin-cor}
Let $\bY = (Y_1,\ldots,Y_d)^\top \sim {\cal SUT}_{d,m}(\0, \bar\bOmega,\bDelta,\btau,\Bar{\bGamma},\nu)$, where $\nu > 2$ and $\bDelta=(\bdelta_1,\dots,\bdelta_m)$. Then, if $m \rightarrow \infty$ with an infinite number of $\bdelta_k \neq \0 \in \R^{d}, k = 1,\dots,m$, the correlation between $Y_i$ and $Y_j$, $\rho_{i,j} \rightarrow \pm 1$ or 0 for $i\neq j$ and $i,j =1, \dots, d$, indicating that $Y_i$ and $Y_j$ exhibit asymptotic linearity or uncorrelation as $m$ increases.  
\end{prop}
\begin{proof}
    To ease the proof, we need to introduce a change of parameterization. In particular,  $\bOmega = \bPsi + \bH\Bar{\bGamma}\bH^\top$ and $\bomega\bDelta = \bH\Bar{\bGamma}$, where $\bPsi$ is also a covariance matrix.
    By Proposition \ref{mean}, $\text{Var}(\bY) = \bomega\{\bDelta\Bar{\bGamma}^{-1}\text{Var}(\bU_*)\Bar{\bGamma}^{-1}\bDelta^\top + \eta(Q_{\bU_*})(\bar{\bOmega} - \bDelta\Bar{\bGamma}\bDelta^\top)\}\bomega$. By re-arranging the terms and plugging in the restrictions,  we can have the following quantity: 
    \begin{align*}
    \text{Var}(\bY) & = \bH\text{Var}(\bU_*)\bH^\top + \eta(Q_{\bU_*})\bPsi 
      = \bH\bL\bL^\top\bH^\top + \eta(Q_{\bU_*})\bPsi, \quad \text{Var}(\bU_*) = \bL\bL^\top \\
     & = \bH_\bL\bH_\bL^\top + \eta(Q_{\bU_*})\bPsi = \sum_{k=1}^m {\textbf{h}_\bL}_k  {\textbf{h}_\bL}_k^\top + \tilde{\bPsi}, \quad \bH_\bL=\bH\bL, \tilde{\bPsi}=\eta(Q_{\bU_*})\bPsi.
   \end{align*}
Now we have that: \\ \vspace{-.1cm}
\begin{align*}
\rho_{i,j} & = \frac{\tilde{\bPsi}_{i,j} + \sum_{k = 1}^m h_{k,i}h_{k,j}} {\sqrt{\tilde{\bPsi}_{i,i} + \sum_{k = 1}^m {h_{k,i}}^2} \sqrt{\tilde{\bPsi}_{j,j} + \sum_{k = 1}^m {h_{k,j}}^2}} \\
& = \frac{\tilde{\bPsi}_{i,j} + \sum_{k = 1}^m h_{k,i}h_{k,j}} {\sqrt{\tilde{\bPsi}_{i,i}\tilde{\bPsi}_{j,j} + \sum_{k=1}^m {h_{k,i}}^2\tilde{\bPsi}_{j,j} + \sum_{k = 1}^m {h_{k,j}}^2}\tilde{\bPsi}_{i,i}+ \sum_{k = 1}^m \sum_{w = 1}^m {h_{k,i}}^2{h_{w,j}}^2},
\end{align*}
where $h_{k,i}$ and $h_{k,j}$ denote the respective $i$-th and $j$-th component of $\textbf{h}_{\bL k}$. In the denominator, we claim that $\sum_{k=1}^m \sum_{w=1}^m {h_{k,i}}^2 {h_{w,j}}^2$ is the dominating term as $m \rightarrow \infty$. Here are the details:
\begin{align*}
    \lim_{m \rightarrow \infty} \frac{\sum_{k=1}^m {h_{k,i}}^2}{\sum_{k=1}^m \sum_{w=1}^m {h_{k,i}}^2 {h_{w,j}}^2} & \leq \lim_{m \rightarrow \infty} \frac{m \cdot \text{max}(\{h_{k,i}\}_{i=1,\dots,m})}{m^2 \cdot \text{min}(\{{h_{k,i}}^2 {h_{w,j}}^2\}_{k,w=1,\dots,m})}=0, \\
    \lim_{m \rightarrow \infty} \frac{\sum_{k=1}^m {h_{k,i}}^2}{\sum_{k=1}^m \sum_{w=1}^m {h_{k,i}}^2 {h_{w,j}}^2} & = 0, \quad \sum_{k=1}^m {h_{k,i}}^2 \geq 0, \quad \sum_{k=1}^m \sum_{w=1}^m {h_{k,i}}^2 {h_{w,j}}^2 \geq 0.
\end{align*}
The same result for comparing with $\sum_{k=1}^m h_{k,j}^2$ can be replicated by replacing $i$ with $j$ in the above derivation. Therefore, we only need to focus on the increments (decrements) $\sum_{k=1}^m h_{k,i}h_{k,j}$ and $\sqrt{\sum_{k=1}^m\sum_{w=1}^m h_{k,i}^2 h_{w,j}^2}$. By the well-known Cauchy-Schwarz inequality, we have 
$
-1 \leq \sum_{k = 1}^m h_{k,i}h_{k,j}/\sqrt{\sum_{k=1}^m\sum_{w=1}^m h_{k,i}^2 h_{w,j}^2} \leq 1
$ 
and the ratio is equal to $\pm 1$ only if $h_{k,i} =  \beta_{{i,j}} h_{k,j}$ with $\beta_{{i,j}} \in \R$, indicating that the $\textbf{h}_{\bL k}$s are pointing to either the same direction or infinitely many times along and against the same unit vector. Note that this situation excludes the directions of the vectors in the planar or hyperplanar subspaces formulated by the main axes, which are special cases that we will explore later. Now, if the rate of the increments (decrements) are equal, then $\lim_{m \rightarrow \infty} \rho_{i,j} = \pm 1$. Otherwise, $\lim_{m \rightarrow \infty} \rho_{i,j} = 0$ because the increment (decrement) rate in the denominator is higher. This is the case when $\textbf{h}_{\bL k}$s point to infinitely many directions.  

Now we assume without loss of generality that $h_{k,i} \neq 0, \forall i \in z \subset \{1,\dots,d\},\text{ and } h_{k,j} = 0, \forall j \in \{1,\dots,d\}-z, \forall k = 1,\dots,m$. Then, all $\textbf{h}_{\bL k} \in \R^{|j|} \subset \R^d$, and:
\begin{align*}
    \lim_{m \rightarrow \infty} \rho_{i,j} & = \lim_{m \rightarrow \infty} 
 \frac{\tilde{\bPsi}_{i,j} + \sum_{k = 1}^m h_{k,i}h_{k,j}} {\sqrt{\tilde{\bPsi}_{i,i} + \sum_{k = 1}^m {h_{k,i}}^2} \sqrt{\tilde{\bPsi}_{j,j} + \sum_{k = 1}^m {h_{k,j}}^2}} 
      = \lim_{m \rightarrow \infty}  \frac{\tilde{\bPsi}_{i,j}} {\sqrt{\tilde{\bPsi}_{i,i} + \sum_{k = 1}^m {h_{k,i}}^2} \sqrt{\tilde{\bPsi}_{j,j}}}  = 0.
\end{align*}
This result has a rather straightforward interpretation: we inflate the variability of $Y_i$s to $\infty$ and leave it unchanged for the remaining $Y_j$s. The $Y_i$s can still exhibit asymptotic linearity or uncorrelation given the above-mentioned choices of ${\textbf{h}_{\bL k}}_{-0} \in \R^{|j|}$, where $-0$ denotes the removal of the 0 terms. 
\end{proof}
By Proposition \ref{lin-cor}, the SUT will show either asymptotic linearity or uncorrelation (symmetry) in the bivariate case as $m$ increases, explaining the humped shapes of Mardia's measures in Figure \ref{mardia}. Per the results, we advise against using excessively large latent dimensions for the SUT. The asymptotic linearity and uncorrelation in the latent dimensions also hold for the SUN distribution because it is a particular case of the SUT distribution. 

Another noteworthy proposition we would like to make is the following.
\begin{prop} \label{SUT-ST}
Let $\bY \sim {\cal SUT}_{d,m}(\bxi,\bOmega,\bDelta,\0,\bar\bGamma,\nu)$. If $\bar\bGamma$ is an equi-correlation matrix with $\bar\bGamma_{i,j} = \rho \approx 1, \forall i \neq j$, then the corresponding $\lim_{\rho \rightarrow 1}\bY \equiv \bY_{ST} \sim {\cal ST}_d$. 
\end{prop}
\begin{proof}
    By the convolution representation in Proposition \ref{convolution}, $\bY = \bxi + \bomega\left(\bDelta\bar\bGamma^{-1}\bU_* + \sqrt{\frac{\nu + Q_{\bU_*}}{\nu + m}}\bW_* \right)$ with $\bU_* = (\bU_0|\bU_0 + \btau > \0)$, where $\bU_0 \sim {\cal T}_m(\0,\bar\bGamma,\nu)$. Now, if $\rho \approx 1$, then $\bU_{*i} \approx \bU_{*j}, \forall i \neq j$, where $i,j=1,\dots,m$, $\bDelta\bar\bGamma^{-1}\bU_* \approx (\sum_{i=1}^m \bH_i)\bU_{*1}$, where $\bH = (\bH_1,\dots,\bH_m) = \bDelta \bar\bGamma^{-1}$. This is exactly the ST distribution, for which the direction of skewness is $\sum_{i=1}^m \bH_i$.
\end{proof}
The same argument can be applied to the SUN and SN distribution. Hence, we recommend to use the SN or ST distribution directly in case of strong latent correlations. 

\section{Non-Identifiability} \label{s9}

\cite{nonke} have demonstrated that the SUN distribution is non-identifiable subject to permutations, ${\cal P}(m) = \{\bP \in \R^{\textit{m} \times \textit{m} }|\bP\bP^\top = \bP^\top\bP = \bI \text{ and } \bP\mathbf{1}_\textit{m}  = \mathbf{1}_\textit{m} \}$, of its latent variables if $m > 1$. In particular, the random vector $\bX \sim \SUN_{d,m}(\bxi,\bOmega,\bDelta,\btau,\bGamma)$ is equal in distribution to the random vector $\bX_p \sim \SUN_{d,m}(\bxi,\bOmega,\bDelta_p,\btau_p, \bGamma_p)$, where $\bDelta_p=\bDelta\bP^\top$, $\btau_p=\bP\btau$, and $\Bar{\bGamma}_p=\bP\Bar{\bGamma}\bP^\top$. 
Therefore, two sets of parameters yield identical probability densities for the same realization. Furthermore, \cite{nonke} have shown that the non-identifiability holds also for the unified skew-elliptical (SUE) class and even more generally for  selection distributions \citep{ABG2006}. Consequently, the SUT is no exception. 

Non-identifiability is problematic, especially for parameter inference, because the resulting optimization curve could possess multiple peaks and mislead the optimization algorithms to erroneous stopping values, rendering the SUT family non-applicable. One possible approach to address this issue includes ranking the components $\tau_i$, $i = 1, \dots, m$, with a strictly increasing (decreasing) order. However, this approach does not cover the case $\btau = \0$. 

Some sub-models discussed in \cite{nonke} can also be applied in the SUT case. For instance, it is possible to eradicate the ordering flexibility of $\bDelta$ by imposing $\bDelta = \delta\bOmega^{1/2}$ or other similar relationships between $\bDelta$ and $\bOmega$.  Another path involves making $\Bar{\bGamma}$ an equi-correlation matrix and $\bDelta$ having identical entries. Therefore, here is a list of some identifiable SUT distributions:
\begin{itemize}
    \item[1)] ${\cal SUT}_{d,m}(\bxi,\bOmega,\bdelta \boldsymbol{1}_m^\top,\tau \boldsymbol{1}_m,(1-\rho)\bI_m + \rho \boldsymbol{1}_m \boldsymbol{1}_m^\top, \nu)$, where $\btau = \tau \boldsymbol{1}_m$ $(\tau\in \R)$, $\bDelta=\bdelta \boldsymbol{1}_m^\top$ $(\bdelta\in\R^d)$ and
$\Bar{\bGamma} = (1-\rho)\bI_m + \rho \boldsymbol{1}_m \boldsymbol{1}_m^\top$, with $\rho\in(-\frac{1}{m-1},1)$. Moreover, by Proposition \ref{SUT-ST}, when $\rho \approx 1$, it is preferable to opt for the ST distribution; 
 \item[2)]${\cal SUT}_{d,m}(\bxi,\bOmega,\bDelta,\alpha \boldsymbol{1}_m + \beta \bj_m,\bar \bGamma, \nu)$, where $\btau = \alpha \boldsymbol{1}_m + \beta \bj_m$ $(\alpha,\beta \in \R, \beta \neq 0)$ and $\bj_m = (1,\ldots,m)^\top$;
 \item[3)] ${\cal SUT}_{d,d}(\bxi,\omega^2 \bar\bOmega, \omega \delta (1 + \delta^2)^{-1/2}\bar\bOmega,\0,\bar\bOmega, \nu)$, where $\omega \in \R$ and $\delta \in \R$;
 \item[4)] ${\cal SUT}_{d,d}(\bxi,\bOmega,\delta\bOmega^{1/2},\0,\bI_d,\nu)$, where $\delta \in \R$.
\end{itemize}
In addition, these sub-models can be combined to formulate various new identifiable cases.

\section{Discussion}\label{s10}

In this work, we conducted a comprehensive exploration of the properties of the SUT distribution. The SUT generalizes the ST distribution proposed in \cite{azzalini2003distributions} so that the latent variables can have specified mean $\btau$ and correlation $\Bar{\bGamma}$. We derived stochastic representations and a SUN-scale mixture method to construct the SUT random variable. Moreover, we described numerous formal probabilistic properties, such as linear transformations, marginals, conditionals, among many others. In addition, the SUT can also be viewed as a generalization of the SUN, which is retrieved by letting $\nu \rightarrow \infty$. We have also provided possible solutions to the non-identifiability associated with the SUT, rendering the distribution applicable in practice. 

 Although the EM algorithm can provide inference for SN parameters, a well-developed inference mechanism for the SUN and the SUT distributions currently needs to be developed. \cite{gupta2012estimation} applied the method of moments on a particular case of the SUN with $\btau = \0$, $\bDelta = \delta\bOmega^{1/2}$, and $\Bar{\bGamma} = \bI_m$ because such detailed specification can significantly simplify the computation of the moments. Nonetheless, the resulting estimates exhibit numerical instability although unbiased. In addition, the method has only been tested up to bivariate data. Therefore, a more general inference scheme is needed.  Only after the successful development of a proper inference algorithm for the SUT distribution can it be applied to real datasets.

\setstretch{1.1}
\bibliographystyle{apalike2}
\bibliography{references}
\end{document}